\documentclass[11pt,a4paper,preprintnumbers,nofootinbib, superscriptaddress]{revtex4}

\usepackage[T1]{fontenc}
\usepackage[utf8]{inputenc}
\usepackage{amsmath}
\usepackage{amssymb}
\usepackage{hyperref}
\usepackage[final]{graphicx}

\usepackage[english]{babel}

\usepackage{graphicx}
\usepackage[dvipsnames]{xcolor}
\usepackage{subfigure}

\usepackage{braket}

\usepackage{amsfonts}
\usepackage{amstext}
\usepackage{slashed}

\usepackage{microtype}

\usepackage{multirow}
\usepackage{booktabs}

\usepackage{placeins}
\usepackage{verbatim}

\DeclareMathOperator{\Tr}{Tr}

\newcommand{\im}{\mathrm{i}}
\newcommand{\dd}{d}

\def\GeV{{\rm GeV}}

\def\A{\mathcal{A}}

\def\L{\mathcal{L}}

\def\O{\mathcal{O}}

\def\V{\mathcal{V}}

\def\ubar{\overline{u}}
\def\dbar{\overline{d}}
\def\sbar{\overline{s}}

\def\qbar{\overline{q}}

\begin{document}

\title{ Radiative three-body $D$-meson decays in and beyond the standard model  }

\author{Nico Adolph}
\email{nico.adolph@tu-dortmund.de}
\affiliation{Fakult\"at Physik, TU Dortmund, Otto-Hahn-Str.4, D-44221 Dortmund, Germany}
\affiliation{Department of Physics, University of Cincinnati, Cincinnati, OH 45221, USA}
\author{Joachim Brod}
\email{joachim.brod@uc.edu}
\affiliation{Department of Physics, University of Cincinnati, Cincinnati, OH 45221, USA}
\author{Gudrun Hiller}
\email{ghiller@physik.uni-dortmund.de}
\affiliation{Fakult\"at Physik, TU Dortmund, Otto-Hahn-Str.4, D-44221 Dortmund, Germany}
\preprint{DO-TH 20/11}

\begin{abstract}

We study radiative charm decays $D \to P_1 P_2 \gamma$,
$P_{1,2}=\pi,K$ in QCD factorization at leading order and within heavy
hadron chiral perturbation theory. Branching ratios including
resonance contributions are around $\sim 10^{-3}$ for the
Cabibbo-favored modes into $K \pi \gamma$ and $\sim 10^{-5}$ for the
singly Cabibbo-suppressed modes into $\pi^+ \pi^- \gamma, K^+ K^-
\gamma$, and thus in reach of the flavor factories BES~III and
Belle~II.  Dalitz plots and forward-backward asymmetries reveal
significant differences between the two QCD frameworks; such
observables are therefore ideally suited for a data-driven
identification of relevant decay mechanisms in the standard-model
dominated $D \to K \pi \gamma$ decays. This increases the potential to
probe new physics with the $D \to \pi^+ \pi^- \gamma$ and $D \to K^+
K^- \gamma$ decays, which are sensitive to enhanced dipole operators.
CP asymmetries are useful to test the SM and look for new physics in
neutral $|\Delta C|=1$ transitions. Cuts in the Dalitz plot enhance
the sensitivity to new physics due to the presence of both $s$- and
$t,u$-channel intermediate resonances.

\end{abstract}

\maketitle

\section{Introduction}

Decays of charmed hadrons provide unique avenues for studying flavor
in the up-quark sector, complementary to $K$ and $B$ physics, and with
great opportunities for experimental study at the
LHCb~\cite{Cerri:2018ypt}, Belle~II~\cite{Kou:2018nap}, and
BES~III~\cite{Ablikim:2019hff} experiments.  We discuss the three-body
Cabibbo-favored standard-model (SM) dominated modes $D \to K \pi
\gamma$ as well as the Cabibbo-supressed modes $D \to \pi \pi \gamma$
and $D \to K K \gamma$. The latter receive $|\Delta C|=1$ flavor
changing neutral current (FCNC) contributions and are sensitive to new
physics (NP).  Our goal is to study QCD and flavor dynamics in and
beyond the standard model (BSM) in the charm sector.  Multi-body
decays supply off-resonant contributions to $D_{(s)} \to V \gamma$,
$V=\rho, \bar K, \phi$~\cite{deBoer:2017que} and, due to their richer
final states, provide opportunities for SM tests through angular
observables, such as polarization studies in $D \to K_1 (\to K \pi
\pi) \gamma$ decays~\cite{Adolph:2018hde}. Due to the poor convergence
of the expansion in inverse powers of the charm-quark mass, $1/m_c$,
strategies to probe for NP in $D$ decays are based on null tests,
exploiting approximate symmetries of the SM, such as CP and flavor
symmetries, or flavor universality~\cite{deBoer:2018buv}.

We perform a comprehensive study of available theory tools for
radiative charm decay amplitudes. A new result is the analysis of $D
\to P_1 P_2 \gamma$ at leading order QCD factorization (QCDF), with
the $P_1 P_2$-form factor as a main ingredient. The framework is
formally applicable for light and energetic $(P_1 - P_2)$ systems. At
the other end of the kinematic spectrum, for large $(P_1 - P_2)$
invariant masses, we employ the soft-photon approximation. We also
re-derive the heavy-hadron chiral perturbation theory (HH$\chi$PT)
amplitudes for $D \to K \pi \gamma$ decays put forward in
Refs.~\cite{Fajfer:2002bq,Fajfer:2002xf}, and provide results for the
FCNC modes $D \rightarrow \pi^+ \pi^- \gamma$ and $D \rightarrow K^+
K^- \gamma$. We find differences between our results and those in
\cite{Fajfer:2002bq} which we detail in Appendix~\ref{app:HQCHPT form
  factors}.

We compare the predictions of the QCD methods, with the goal to
validate and improve the theoretical description via the study of the
SM dominated decays. Then, we work out the NP sensitivities of the
FCNC modes $D \to \pi \pi \gamma$ and $D \to K K \gamma$ in several
distributions and observables.

The methods we employ, such as QCDF, are well-known and established
methods in $B$ physics. In charm physics the expansion parameters are
numerically larger, and the systematic computation of amplitudes from
first principles becomes a challenging task -- hence the importance of
null tests.  On the other hand, while $B$ physics has entered the
precision era, very few radiative or semileptonic rare charm decays
have been observed so far. Notably, there are no data on $D \to PP
\gamma$ decay rates or its distributions. Therefore, while QCDF and
HH$\chi$PT are not expected to perform as well as in $B$ physics, we
take their qualitative agreement within their ranges of validity as
indicative of providing the correct order of magnitude in charm
physics. This is sufficient to make progress given the experimental
situation and leaves room for theory improvements, which can come also
in a data-driven way, as we very concretely propose to do using decay
distributions.

The paper is organized as follows: In Section~\ref{sec:QCD} we
introduce kinematics and distributions, and use QCD factorization
methods (Section~\ref{sec:QCDF}) and Low's theorem
(Section~\ref{sec:low}) for predictions for small and large
$PP$-invariant masses, respectively. In Section~\ref{sec:HHCHPT} we
work out the HH$\chi$PT amplitudes and Dalitz plots. We provide SM
predictions for branching ratios and the forward-backward asymmetries
in all three approaches and compare them in Section~\ref{sec:compare}.
In Section~\ref{sec:BSM} we analyze the maximal impact of BSM
contributions on the differential branching ratios and the
forward-backward asymmetries. New-physics signals in CP asymmetries
are worked out in Section~\ref{sec:CP}. We conclude in
Section~\ref{sec:con}. Auxiliary information on parametric input
parameters and form factors is provided in two appendices.

\section{Radiative three-body decays in QCD frameworks \label{sec:QCD}}

We review the kinematics of the radiative three-body decays $D \to P_1
P_2 \gamma$ in section~\ref{sec:kin}.  We then work out the SM
predictions using QCD factorization methods in section~\ref{sec:QCDF},
Low's theorem in section~\ref{sec:low}, and HH$\chi$PT in
section~\ref{sec:HHCHPT}.

\subsection{Kinematics \label{sec:kin}}

The general Lorentz decomposition of the $D(P) \rightarrow P_1(p_1)
P_2(p_2) \gamma(k, \epsilon^*)$ amplitude reads
\begin{align} \label{eq:mainA}
  \A(D \rightarrow P_1 P_2 \gamma) = A_-(s,t)  \left[(p_1 \cdot k)(p_2 \cdot  \epsilon^{*}) - (p_2 \cdot  k)(p_1 \cdot  \epsilon^{*})\right] + A_+(s,t) \epsilon^{\mu\alpha\beta\gamma}
   \epsilon^{*}_\mu p_{1\alpha} p_{2\beta} k_\gamma \,,
\end{align}
with parity-even $(A_+)$ and parity-odd $(A_-)$ contributions.  The
four-momenta of the $D$, $P_1$, $P_2$ and photon are denoted by $P,
p_1,p_2$ and $k$, respectively; the photon's polarization vector is
$\epsilon^*$.
Above, $s=(p_1 + p_2)^2$ and $t=(p_2 + k)^2$ refer to the squared
invariant masses of the $P_1$--$P_2$ and $P_2 $--$\gamma$ systems,
respectively.  We denote the negatively charged meson or the
$\overline{K}^0$ by $P_2$. Moreover, $\epsilon^{\mu\alpha\beta\gamma}$
is the totally antisymmetric Levi-Civita tensor; we use the convention
$\epsilon^{0123}=+1$. The double differential decay rate is then given
by
\begin{align}
  \frac{\dd{}^2\Gamma (D \to P_1 P_2 \gamma)}{\dd{}s \dd{}t} = \frac{1}{32(2\pi)^3 m_D^3} \left(|\A_L|^2 + |\A_R|^2\right)\, ,
\end{align}
where $m_D$ is the $D$-meson mass. We obtain
\begin{equation}
\begin{split}
 \frac{\dd{}^2\Gamma}{\dd{}s \dd{}t} & = \frac{ |A_-|^2 + |A_+|^2}{128 (2\pi)^3 m_D^3}\\
 & \quad \times \big[m_1^2(t-m_2^2)(s-m_D^2)-m_2^4m_D^2 - st(s+t-m_D^2) + m_2^2(st+(s+t)m_D^2 - m_D^4)\big]\,.
\end{split}
\end{equation}
The subscript $L(R)$ refers to the left- (right-)handed polarization
state of the photon, and
\begin{align}
 \A_L &= \frac{1}{\sqrt{2}}(A_- + \im A_+)  x  \, , \quad \quad 
 \A_R = \frac{1}{\sqrt{2}}(A_- - \im A_+) x  \, , \\
 x&= \sqrt{m_1^2(t-m_2^2)(s-m_D^2)-m_2^4m_D^2 - st(s+t-m_D^2) + m_2^2(st+(s+t)m_D^2 - m_D^4)}/2 \, ,
\end{align}
where $m_1 (m_2)$ denotes the mass of the $P_1 (P_2)$ meson. The
single differential distribution in the squared invariant di-meson
mass is then given by
\begin{align}
  \begin{split}
    &\frac{\dd{}\Gamma}{\dd{}s} = \int_{t_{\rm min}}^{t_{\rm max}} dt  \frac{\dd{}^2\Gamma}{\dd{}s \dd{}t} \, , \\
    &t_{\rm min}= \frac{(m_D^2 - m_{1}^2 + m_{2}^2)^2}{4s}  - \left(\sqrt{\frac{(s - m_{1}^2 + m_{2}^2)^2}{4s} - m_2^2} + \frac{m_D^2 - s}{2\sqrt{s}}\right)^2\, ,\\  
    &t_{\rm max}= \frac{(m_D^2 - m_{1}^2 + m_{2}^2)^2}{4s}  - \left(\sqrt{\frac{(s - m_{1}^2 + m_{2}^2)^2}{4s} - m_2^2} - \frac{m_D^2 - s}{2\sqrt{s}}\right)^2 \, ,
  \end{split}
\end{align}
and $(m_{1}+m_{2})^2  \leq s \leq m_D^2$.

\subsection{QCD Factorization}\label{sec:QCDF}

Rare $c \rightarrow u \gamma$ processes can be described by the
effective four-flavor Lagrangian~\cite{deBoer:2017que}
\begin{align}
    \L_{\text{eff}} = \frac{4G_F}{\sqrt{2}}\left(\sum_{q, q'\in \{d, s\}}V_{cq}^* V_{uq'}\sum_{i=1}^2 C_i O_i^{(q, q')}  +\sum_{i=3}^6 C_i O_i + \sum_{i=7}^8 \left(C_i O_i + C_i^\prime O_i^\prime\right)\right)\, .
\end{align}
Here, $G_F$ is Fermi's constant and $V_{ij}$ are elements of the
Cabibbo-Kobayashi-Maskawa (CKM) matrix. The operators relevant to this
work are given by
\begin{equation}
  \begin{alignedat}{2}
    &O_1^{(q, q')} = \left(\ubar_L \gamma_\mu T^a q_L'\right) \left(\qbar_L \gamma^\mu T^a c_L\right)\, , \quad &&O_2^{(q, q')} = \left(\ubar_L \gamma_\mu q_L'\right) \left(\qbar_L \gamma^\mu c_L\right)\, ,\\
    &O_7 = \frac{e m_c}{16\pi^2}  \left(\ubar_L \sigma^{\mu \nu} c_R\right)F_{\mu \nu}\, , &&O_7^\prime = \frac{e m_c}{16\pi^2}  \left(\ubar_R \sigma^{\mu \nu} c_L\right)F_{\mu \nu}\, ,
  \end{alignedat}
\end{equation}
where the subscripts $L(R)$ denote left-(right-)handed quark fields,
$F_{\mu \nu}$ is the photon field strength tensor, and $T^a$ are
generators of $SU(3)$ normalized to $\text{Tr}\{T^aT^b\} =
\delta^{ab}/2$, respectively. Because of an efficient cancellation due
to the Glashow-Iliopoulos-Maiani mechanism, only the four-quark
operators $O_{1,2}^{(q, q')}$ are induced at the $W$-scale $\mu_W$ and
receive order-one coefficients at the scale $\mu_c \sim m_c$ of the
order of the charm-quark mass. At leading order in the strong coupling
$\alpha_s$, the coefficients are given for $\mu_c \in
\left[m_c/\sqrt{2}, \sqrt{2}m_c\right]$ by~\cite{deBoer:2017que}
\begin{align} \label{eq:Cbar}
    &C_1 \in \left[-1.28, -0.83\right], \qquad C_2 \in \left[1.14, 1.06\right]\, ,
    &\tilde{C}\equiv\frac{4}{9}C_1 + \frac{1}{3}C_2 \in \left[-0.189, -0.018\right]\, .  
\end{align} 
The peculiar combination of Wilson coefficients $\tilde{C}$ arises in
the weak annihilation amplitude (see below); note that an accidental
numerical cancellation occurs in this combination, leading to a large
scale uncertainty (see Table~\ref{tbl:branching_ratios}). This effect
is partially mitigated by higher-order QCD corrections which we do not
take into account in this work; see, e.g., Ref.~\cite{deBoer:2017que}.
The tiny SM contributions to $C_{3-8}$ are a result of renormalization
group running and finite threshold corrections at the bottom-mass
scale, and can be neglected for the purpose of this work. For
instance, the SM contribution of the electromagnetic dipole operator
$O_7$ is strongly suppressed, $|C_7^\text{eff}| \simeq \O(0.001)$ at
$\mu_c=m_c$ at next-to-next-to-leading order~\cite{deBoer:2018buv}.

\begin{figure}
  \centering
  \includegraphics[width=0.6\linewidth]{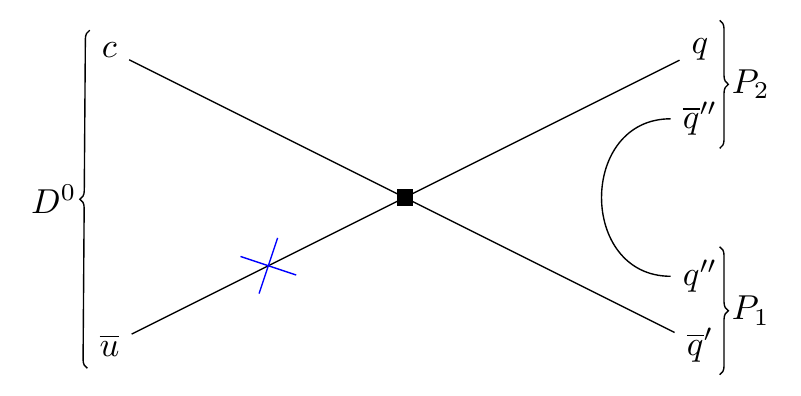}
  \caption{The weak annihilation diagram for $D \rightarrow P_1 P_2
    \gamma$. The blue cross indicates the dominant photon emission
    from the light quark of the $D$ meson. Photon emission from the
    other quark lines is suppressed by powers of
    $\lambda_\text{QCD}/m_c$ or $\alpha_s$.}
  \label{fig:WAcontribution}
\end{figure}

In this section we use QCDF methods~\cite{Beneke:2000ry, Bosch:2001gv,
  DescotesGenon:2002mw} to calculate the leading weak annihilation
(WA) contribution shown in Fig.~\ref{fig:WAcontribution}.
We obtain
\begin{align}
  \begin{split}
    &\A^{\text{WA}}_- = i \frac{G_F e}{\sqrt{2}} \tilde{C}  \frac{f_D Q_u}{\lambda_D (v\cdot k)} \sum_{q, q'\in \{d, s\}} V_{cq}^* V_{uq'}  f^{P_1 P_2}_{(q, q')}(s) \, ,\\
    &\A^{\text{WA}}_+ = \frac{G_F e}{\sqrt{2}} \tilde{C}  \frac{f_D Q_u}{\lambda_D (v\cdot k)} \sum_{q, q'\in \{d, s\}} V_{cq}^* V_{uq'}  f^{P_1 P_2}_{(q, q')}(s)  \, ,  \label{eq:QCDF-Amplitude}
  \end{split}
\end{align}
where $Q_u=2/3$ denotes the electric charge of the up-type quarks, and
we decomposed $P=v\,m_D$. The nonperturbative parameter $\lambda_D
\sim \Lambda_{\text{QCD}}$ is poorly known and thus source of large
theoretical uncertainties. In the following we use $\lambda_D = 0.1\,
\GeV$ \cite{deBoer:2017que}. For the final states $\pi^+ \pi^- \gamma$
and $K^+ K^- \gamma$, the remaining form factors $f^{P_1 P_2}_{(q,
  q')}(s)$ can be expressed in terms of the electromagnetic pion and
kaon form factors~\cite{Bruch:2004py}. For the final states $ \pi^+
K^-\gamma$ and $\pi^0 \overline{K} \gamma$, we use the form factors
extracted from $\tau^- \rightarrow \nu_\tau K_s \pi^0$ decays
\cite{Boito:2008fq} in combination with isospin relations. We obtain
for the non-vanishing form factors
\begin{align}  \label{eq:iso}
  \begin{split}
    &f^{\pi^+ \pi^-}_{(d,d)}(s) = - F^{\text{em}}(s)\, ,\\
    &f^{K^+K^-}_{(d,d)}(s) = 3F^{(I=0)}_{K^+}(s) - F^{(I=1)}_{K^+}(s)\, ,\\
    &f^{K^+ K^-}_{(s,s)}(s) = -3F^s_{K^+}(s)\,  ,\\
    &f^{\pi^+K^-}_{(s,d)}(s)  = -f^{\overline{K}\pi^-}_+(s)\, , \\
    &f^{\pi^0\overline{K}^0}_{(s,d)}(s)  = \frac{1}{\sqrt{2}}f^{\overline{K}\pi^-}_+(s) \, .
  \end{split}
\end{align}
More details about the form factors are given in
appendix~\ref{appendix:em_weak_form_factors}. We recall that QCDF
holds for light and energetic $P_1$--$P_2$ systems. This limits the
validity of the results to $s \lesssim 1.5\, \GeV^2$, corresponding to
an approximate upper limit on a light hadron's or hadronic system's
invariant mass squared, including the $\phi$. The WA decay amplitudes
are independent of $t$.

\subsection{ Soft photon approximation \label{sec:low}}

Complementary to QCDF, we use Low's theorem~\cite{Low:1958} to
estimate the decay amplitudes in the limit of soft photons. This
approach holds for photon energies below
$m_P^2/E_P$~\cite{DelDuca:1990}, which results in $s\gtrsim 2.3 \,
\GeV^2$ for $D \rightarrow K^+ K^- \gamma$ and $s\gtrsim 3.4 \,
\GeV^2$ for decays with a final-state pion. The amplitude is then
given by~\cite{Cappiello:2012vg}
\begin{align}
  \A_-^{\rm Low}=-\frac{e\A(D \to P_1 P_2)}{(p_1\cdot k)(p_2\cdot k)}\, , \label{eq:Low_theorem}
\end{align}
while $ \A_+^{\rm Low}=0$. There is no such contribution to $D \to
\pi^0 \bar K^0 \gamma$, since only neutral mesons are involved.  The
modulus of the $D \rightarrow P_1 P_2$ amplitudes can be extracted
from branching ratio data using
\begin{align}
  \mathcal{B}(D \rightarrow P_1 P_2) = \frac{|\A(D \rightarrow P_1 P_2)|^2}{16\pi m_D \Gamma_D}\sqrt{\left(1-\frac{(m_1 + m_2)^2}{m_D^2}\right)\left(1-\frac{(m_1 - m_2)^2}{m_D^2}\right)}\, ,
\end{align}
where $\Gamma_D$ is the total width of the D meson. Using the
parameters given in appendix \ref{app:Parameter}, we obtain
\begin{align}
  \begin{split}
    &\left| \A(D \rightarrow \pi^+ \pi^-)\right| = (4.62 \pm 0.04) \cdot 10^{-7}\, \text{GeV}\, , \\
    &\left| \A(D \rightarrow \pi^+ K^-)\right| = (2.519 \pm 0.014) \cdot 10^{-6}\, \text{GeV}\, , \\
    &\left| \A(D \rightarrow K^+ K^-)\right| = (8.38 \pm 0.09) \cdot 10^{-7}\, \text{GeV}\, .
  \end{split}
\end{align}
Low's theorem predicts that the differential decay rate behaves as \cite{DAmbrosio:1994bks}
\begin{align}
  \frac{\dd{}\Gamma}{\dd{}s} \sim \frac{1}{m_D^2-s}\, .
\end{align}
Consequently, there is a singularity at the boundary of the phase
space. This corresponds to a vanishing photon energy in the $D$
meson's rest frame. The tail of the singularity dominates the decay
rate for small photon energies.  We remove these events for integrated
rates by cuts in the photon energy, as they are of known SM origin and
hamper access to flavor and BSM dynamics.

\subsection{HH$\chi$PT}
\label{sec:HHCHPT}

As a third theory description we use the framework of heavy hadron
chiral perturbation theory (HH$\chi$PT), which contains both the heavy
quark and the $SU(3)_L \times SU(3)_R$ chiral symmetry. The effective
Lagrangian was introduced in \cite{Wise:1992,Burdman:1992,Yan:1992}
and extended by light vector resonances by Casalbuoni \textit{et al.}
\cite{Casalbuoni:1993}. We follow the approach of Fajfer \textit{et
  al.}, who studied radiative two-body decays $D\rightarrow V \gamma$
\cite{Bajc:1994ui,Fajfer:1998dv} and Cabibbo allowed three-body decays
$D\rightarrow K^- \pi^+ \gamma$ \cite{Fajfer:2002bq} and $D
\rightarrow \overline{K}^0 \pi^0 \gamma$ \cite{Fajfer:2002xf} in this
way.\\ The light mesons are described by $3\times3$ matrices
\begin{align}
  &u = \exp\left(\frac{\mathrm{i} \Pi}{f}\right)\, , \qquad\Pi = \left( \begin{array}{ccc}
    \frac{\pi^0}{\sqrt2} + \frac{\eta_8}{\sqrt6} & \pi^+ & K^+  \\
    \pi^- & -\frac{\pi^0}{\sqrt2} + \frac{\eta_8}{\sqrt6} & K^0  \\
    K^- & \overline{K}^0 &  -\frac{2\eta_8}{\sqrt6}  \\
    \end{array}\right)\, , \\
  &\hat{\rho}_\mu = \mathrm{i} \frac{g_v}{\sqrt{2}}\rho_\mu\, , \qquad \rho_\mu = \left( \begin{array}{ccc}
      \frac{\rho_\mu^0 + \omega_\mu}{\sqrt2} & \rho_\mu^+ & K^{\star+}_\mu  \\
      \rho_\mu^- & \frac{-\rho_\mu^0 + \omega_\mu}{\sqrt2} & K^{\star0}_\mu  \\
      K^{\star-}_\mu & \overline{K} ^{\star0}_\mu &  \Phi_\mu  \\
      \end{array}\right)\, ,
\end{align}
where $f\simeq f_\pi$ is the pion decay constant and $g_v =
5.9$~\cite{Bando:1985}. To write down the photon interaction with the
light mesons in a simple way, we define two currents
\begin{align}
  \begin{split}
    \V_\mu &= \frac{1}{2}\left(u^\dagger D_\mu u + u D_\mu u^\dagger \right) \, , \\
    \A_\mu &= \frac{1}{2}\left(u^\dagger D_\mu u - u D_\mu u^\dagger \right) \, .
  \end{split} 
\end{align}
Here, the covariant derivative acting on $u$ and $u^\dagger$ is given
by $D_\mu u^{(\dagger)} = \partial_\mu u^{(\dagger)} + \im e B_\mu Q
u^{(\dagger)}$, with the photon field $B_\mu$ and the diagonal charge
matrix $Q=\text{diag}(2/3, -1/3, -1/3)$. The even-parity strong
Lagrangian for light mesons is then given by~\cite{Bando:1985}
\begin{equation}
  \L_{\text{light}} = -\frac{f^2}{2} \left[ \Tr_F\left( \A_\mu \A^\mu \right) + a \Tr_F \left( \left( \V_\mu - \hat{\rho}_\mu\right)^2 \right)\right] + \frac{1}{2g_v^2}\Tr_F \left( F_{\mu \nu}(\hat{\rho}) F^{\mu \nu}(\hat{\rho})\right)\, ,
\end{equation}
where $F_{\mu \nu}(\hat{\rho}) = \partial_\mu \hat{\rho}_\nu -
\partial_\nu \hat{\rho}_\mu + \left[\hat{\rho}_\mu, \hat{\rho}_\nu
  \right]$ denotes the field strength tensor of the vector
resonances. In general, $a$ is a free parameter, which satisfies $a=2$
in case of exact vector meson dominance (VMD). In VMD there is no
direct vertex that connects two pseudoscalars and a photon. In this
case, the photon couples to pseudoscalars via a virtual vector meson.
Analogously, the matrix element $\braket{P_1 P_2| \qbar \gamma^\mu
  (1-\gamma_5)q^\prime|0}$ also vanishes. However, we do not use the
case of VMD and exact flavor symmetry, but allow for $SU(3)$ breaking
effects. Therefore, we choose to set $a=1$ and replace the model
coupling $g_v$, decay constant $f$, and vector meson mass
$m_V=\sqrt{a/2}g_v f$ in $\L_\text{light}$ with the respective
measured masses, decay constants and couplings $g_v =
\sqrt{2}m_V^2/g_V$. They are defined by
\begin{align}
  \braket{V(q, \eta)|j_V^\mu|0} = \eta^{*\mu}(q) g_V(q^2)\, ,
\end{align}
where $j_{K^\star, \overline{K}^\star, K^{\star\pm}, \Phi}^\mu = \qbar
\gamma^\mu q^\prime$ and $j_{\omega,
  \rho}^\mu=\frac{1}{\sqrt{2}}(\ubar \gamma^\mu u \pm \dbar \gamma^\mu
d)$. Here, $q$ and $\eta$ denote the vector meson's momentum and
polarization vector, respectively. For our numerical evaluation we use
$g_V(0) \simeq g_V(m_V^2)=m_Vf_V$, where $f_V$ is the vector meson
decay constant with mass dimension one. With these couplings the
following $V\gamma$ interactions arise~\cite{Fajfer:1998dv}
\begin{equation}
  \L_{V_0\gamma} = -\frac{e}{\sqrt{2}} B_\mu \left( g_\rho \rho^{0\mu} + \frac{1}{3} g_\omega \omega^\mu - \frac{\sqrt{2}}{3} g_\Phi \Phi^\mu \right)\, . \label{eq:L_Vgamma}
\end{equation}
Instead of the VVP interactions generated by the odd-parity Lagrangian
\cite{Bramon:1995}, we use effective VP$\gamma$ interactions
\begin{align}
  \L_{VP\gamma} = - \frac{1}{2}e g_{VP\gamma} \epsilon_{\mu \nu \rho \sigma} F^{\mu \nu}(B) \partial^\rho V^\sigma P^\dagger + \text{h.c.} \label{eq:L_VPgamma}
\end{align}
and determine the effective coefficients $g_{VP\gamma}$ from
experimental data~\cite{Fajfer:2002bq, Fajfer:1997bh}
\begin{align}
  \Gamma(V\rightarrow P\gamma) = \frac{\alpha_{\text{em}} m_V^3}{24} |g_{VP\gamma}|^2 \left(1-\frac{m_P^2}{m_V^2}\right)^3\, .
\end{align}

The heavy pseudoscalar and vector mesons are represented by $4\times
4$ matrices
\begin{align}
  \begin{split}
    H_a &= \frac{1}{2}\left(1+\slashed{v}\right)\left( P^\star_{a\mu} \gamma^\mu - P_a \gamma_5 \right)\, ,\\
  \overline{H}_a &= \gamma^0 H_a^\dagger \gamma^0 = \left( P^{\star\dagger}_{a\mu} \gamma^\mu + P^\dagger_a \gamma_5 \right)\frac{1}{2}\left(1+\slashed{v}\right)\, ,
  \end{split} 
\end{align}
where $P^{\star(\dagger)}_{a\mu}$, $P^{(\dagger)}_a$ annihilate
(create) a heavy spin-one and spin-zero meson $h_a$ with quark flavor
content $c{\overline{q}_a}$ and velocity $v$, respectively. The
annihilation operators are normalized as
\begin{align}
  \begin{split}
    &\braket{0 | P_a | h_a(v)} = 1\, , \\
    &\braket{0 | P_a^{\star \mu} | h_a^\star(v, \eta)} = \eta^\mu \, .
  \end{split}
\end{align}
The heavy-meson Lagrangian reads
\begin{align}
  \begin{split}
    \L_{\text{heavy}} &= \im \Tr_D\left( H_a v_\mu \left(D^\mu\right)_{ab} \overline{H}_b\right) + \im g \Tr_D \left( H_a \gamma_\mu \gamma_5 (\A^\mu)_{ab} \overline{H}_b \right) \\
    & \quad + \im \tilde{\beta} \Tr_D \left( H_a v_\mu \left( \V^\mu - \hat{\rho}^\mu \right)_{ab} \overline{H}_b \right)\, ,
  \end{split}
\end{align}
where the covariant derivative is defined as $\left(D^\mu\right)_{ab}
\overline{H}_b = \partial^\mu \overline{H}_a +
\left(\V^\mu\right)_{ab} \overline{H}_b - \im e Q_c B^\mu
\overline{H}_a$, with the electric charge of the charm quark
$Q_c=2/3$. The parameter $g=0.59$ was determined by experimental data
of strong $D^\star \rightarrow D \pi$ decays~\cite{Singer:1999ak,
  Anastassov:2001cw}. The coupling $\tilde{\beta}$ seems to be very
small and will be neglected~\cite{Bajc:1997ey}. The odd-parity
Lagrangian for the heavy mesons is given by
\begin{equation}
  \L = \im \lambda \Tr \left( H_a \sigma_{\mu \nu} F^{\mu \nu}(\hat{\rho})_{ab} \overline{H}_b\right)-\lambda^\prime e \Tr\left( H_a \sigma_{\mu \nu} F^{\mu \nu}(B) \overline{H}_a \right)\, ,
\end{equation}
with $\sigma_{\mu \nu} = \frac{\mathrm{i}}{2}\left[\gamma_\mu,
  \gamma_\nu\right]$. The couplings $\lambda$ and $\lambda^\prime$ can
be extracted from rations $R_\gamma^{0/+}=
\Gamma(D^{\star0/+}\rightarrow
D^{0/+}\gamma)/\Gamma(D^{\star0/+}\rightarrow
D^{0/+}\pi)$. $\lambda=-0.49\, \GeV^{-1}$ and $\lambda^\prime=-0.102\,
\GeV^{-1}$ are in good agreement with data \cite{Fajfer:2002bq}.  The
partonic weak currents can be expressed in terms of chiral currents
as~\cite{Bajc:1995km, Bajc:1994ui}
\begin{align}
  \begin{split}
    (\qbar_a Q)^\mu_{\text{V-A}} \simeq J_{Q\overline{q}_a}^\mu &= \frac{1}{2}\im \alpha \Tr\left( \gamma^\mu (1-\gamma_5) H_b u_{ba}^\dagger \right) + \alpha_1 \Tr \left( \gamma_5 H_b \left(\hat{\rho}^\mu - \V^\mu\right)_{bc} u_{ca}^\dagger \right) \\
    &+ \alpha_2 \Tr \left( \gamma^\mu \gamma_5 H_b v_\alpha (\hat{\rho}^\alpha - \V^\alpha)_{bc} u_{ca}^\dagger \right) + \ldots\, ,\\
    (\qbar_j q_i)_{\text{V-A}}^\mu \simeq J_{ij}^\mu &= \im f^2 \left\{ u \left[ \A^\mu + a \left( \V^\mu - \hat{\rho}^\mu \right) \right] u^\dagger \right\}_{ij}\, ,
  \end{split}
\end{align}
where the ellipsis denotes higher-order terms in the chiral and
heavy-quark expansions. The definition of the heavy-meson decay
constants implies $\alpha=f_h \sqrt{m_h}$. The parameters $\alpha_1$
and $\alpha_2$ can be extracted from $D\rightarrow V$ transition form
factors \cite{Fajfer:2002bq}
\begin{align}
  \begin{split}
    A_1(q^2_{\text{max}}) = 2 \frac{\sqrt{m_D}}{m_D + m_V} \frac{m_V^2}{g_V}\alpha_1\, , \qquad A_2(q^2_{\text{max}}) = 2 \frac{m_D + m_V}{m_D^{\frac{3}{2}}} \frac{m_V^2}{g_V}\alpha_2\, . \label{eq:alpha1/2}
  \end{split}
\end{align}
Using the $D \to K^\star$ form factors \cite{Verma:2011yw} we obtain
$\alpha_1 = 0.188\, \GeV^{\frac{1}{2}}$ and $\alpha_2 = 0.086\,
\GeV^{\frac{1}{2}} $. The signs in \eqref{eq:alpha1/2} are due to the
conventions in \cite{Verma:2011yw}. The weak tensor current is given
by \cite{Casalbuoni:1993nh}
\begin{align}
  \begin{split}
    & \qbar \sigma^{\mu \nu}(1+\gamma_5)Q \simeq J_{Q\overline{q}_a}^{\mu \nu} \\ &  = \frac{1}{2}\im \alpha \Tr\left( \sigma^{\mu \nu} (1+\gamma_5) H_b u_{ba}^\dagger \right) \\
    & \quad + \im \alpha_1 \left(g^{\mu \alpha}g^{\nu \beta} - \frac{1}{2}i \epsilon^{\mu \nu \alpha \beta}\right)\Tr \left( \gamma_5 H_b \left[\gamma_\alpha\left(\hat{\rho}_\beta - \V_\beta\right)_{bc} - \gamma_\beta\left(\hat{\rho}_\alpha - \V_\alpha\right)_{bc}\right]u_{ca}^\dagger \right) \\
    & \quad - \alpha_2 \Tr\left(\sigma^{\mu\nu} \gamma_5 H_b v_\alpha \left(\hat{\rho}^\alpha - \V^\alpha\right)_{bc} u^\dagger_{ca}\right) + \ldots\, ,
  \end{split}
\end{align}
where, again, the ellipsis denotes higher-order terms in the chiral
and heavy-quark expansions.

The parity-even and parity-odd amplitudes are given in terms of four
form factors
\begin{align} \label{eq:HHA}
  \begin{split}
    &A_-^{\rm HH\chi PT} = \frac{G_F e}{\sqrt{2}} \sum_{q, q'\in \{d, s\}}V_{cq}^*V_{uq'}  \left[(C_2 - \frac{1}{6}C_1) \sum_{i} A_i^{(q, q')} + \frac{1}{2}C_1 \sum_{i} E_i^{(q, q')}\right]\, ,\\
    &A_+^{\rm HH\chi PT} = \frac{G_F e}{\sqrt{2}} \sum_{q, q'\in \{d, s\}}V_{cq}^*V_{uq'}  \left[(C_2 - \frac{1}{6}C_1) \sum_{i} B_i^{(q, q')} + \frac{1}{2}C_1 \sum_{i} D_i^{(q, q')}\right]\, .
  \end{split}
\end{align}
\begin{figure}
  \centering
  \includegraphics[width=0.9\linewidth]{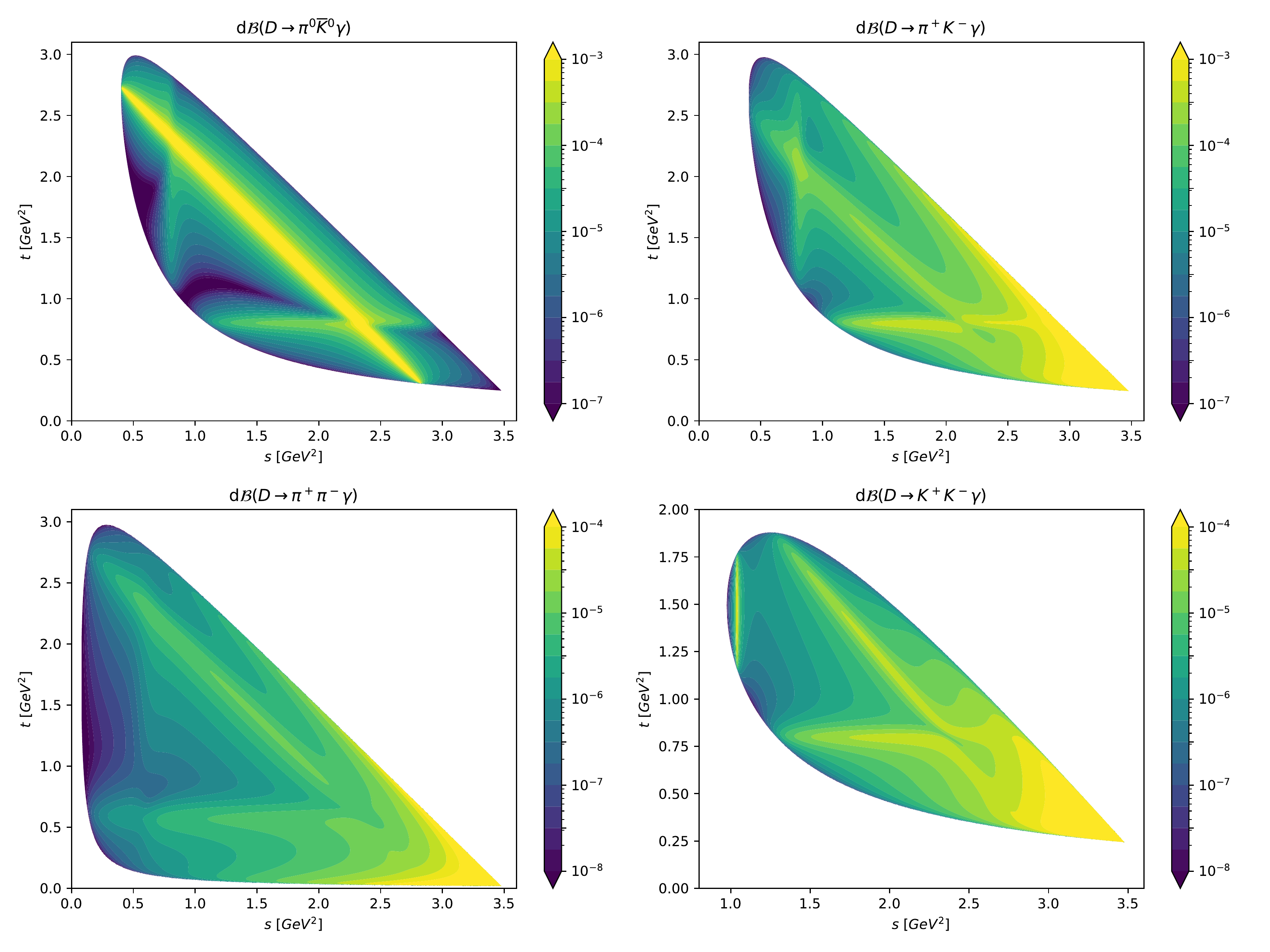}
  \caption{SM Dalitz plots for the decays $D \to \overline{K}^0 \pi^0
    \gamma$ (upper left), $D \to K^- \pi^+ \gamma$ (upper right), $D
    \to \pi^+ \pi^- \gamma$ (lower left) and $D \to K^+ K^- \gamma$
    (lower right) based on HH$\chi$PT at $\mu_c = m_c$. }
  \label{fig:SM_d2BR}
\end{figure}
Here, $A$ and $B$ belong to the charged current operator $(\ubar
q^\prime)_{\mu V-A}(\qbar c)^\mu_{V-A} \equiv 4 O_2^{(q',q)}$ and D
and E to the neutral current operator $(\qbar q^\prime)_{\mu
  V-A}(\ubar c)^\mu_{V-A} \equiv 8 O_1^{(q',q)} + 4 O_2^{(q',q)}
/3$. The corresponding diagrams are shown in
Fig.~\ref{fig:Diagramme_AE} and \ref{fig:Diagramme_BD}. The non-zero
contributions are listed in Appendix~\ref{app:HQCHPT form factors},
where we also provide a list with differences between our results and
those in Ref.~\cite{Fajfer:2002bq}. We neglect the masses of the light
mesons in the form factors, but consider them in the phase space. To
enforce Low's theorem, we remove the bremsstrahlung contributions
$A_{1,2}$ in (\ref{eq:HHA}) and add (\ref{eq:Low_theorem}) to
$A_-^{\rm HH\chi PT}$. For the strong phase we have taken the value
predicted by HH$\chi$PT. In Fig.~\ref{fig:SM_d2BR} we show Dalitz
plots based on the SM HH$\chi$PT predictions. Besides the dominant
bremsstrahlung effects for large s, the intermediate $\rho$, $\omega$,
$K^\star$ and $\Phi$ resonances are clearly visible as bands in $s, t$
and the third Mandelstam variable, $u=(p_1+k)^2=m_D^2+m_1^2
+m_2^2-s-t$.

\section{Comparison of QCD frameworks \label{sec:compare}}

In this section, we compare the predictions obtained using the
different QCD methods in Section~\ref{sec:QCD}. We anticipate
quantitative and qualitative differences between QCDF to leading order
and HH$\chi$PT.
First, we study differential and integrated branching ratios in
Section~\ref{sec:BR}.  In Section~\ref{sec:AFB} we propose to utilze a
forward-backward asymmetry, defined below in Eq.~\eqref{eq:AFB}, to
help disentangling the resonance contributions to the branching
ratios. This subsequently improves the NP sensitivity of the $D \to
P^+ P^- \gamma$ decays.  We consider the U-spin link, exploited
already for polarization-asymmetries in radiative charm
decays~\cite{deBoer:2018zhz}, in Section~\ref{sec:U}.

\subsection{Branching ratios \label{sec:BR}}

\begin{figure}
  \centering
  \includegraphics[width=0.9\linewidth]{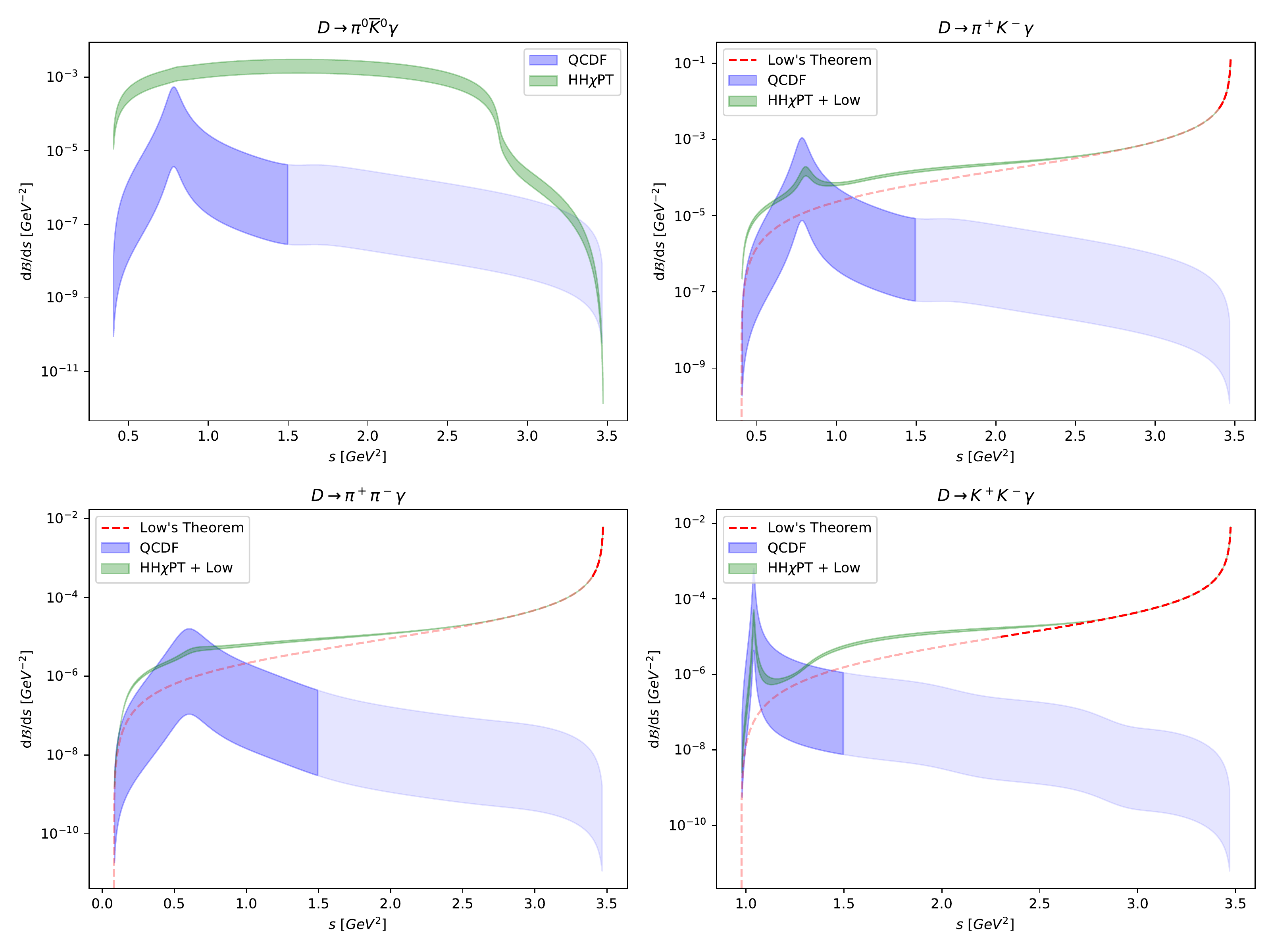}
  \caption{The SM predictions for the differential branching ratios of
    the decays $D \to \overline{K}^0 \pi^0 \gamma$ (upper left), $D
    \to K^- \pi^+ \gamma$ (upper right), $D \to \pi^+ \pi^- \gamma$
    (lower left) and $D \to K^+ K^- \gamma$ (lower right). Blue bands
    correspond to QCDF, green bands to HH$\chi$PT and the red dashed
    lines to the soft photon approximation.  The lighter shaded areas
    and lines illustrate extrapolations beyond the model's region of
    applicability.  QCDF branching ratios are obtained for
    $\lambda_D=0.1\, \GeV$ and are $\propto (0.1\,
    \GeV/\lambda_D)^2$.}
  \label{fig:SM_dBr_Vergleich}
\end{figure}

The branching ratios for the various decay modes, obtained from QCDF
(blue bands), HH$\chi$PT (green bands) and Low's theorem (red dashed
lines), are shown in Fig.~\ref{fig:SM_dBr_Vergleich}.  The width of
the bands represents the theoretical uncertainty due to the $\mu_c$
dependence of the Wilson coefficients.

The shape of the QCDF results is mainly given by the $P_1-P_2$ form
factors and their resonance structure. For the $D \to P_1^+ P_2^-
\gamma$ decays, the high-$s$ regions of the HH$\chi$PT predictions are
dominated by bremsstrahlung effects. Since we have replaced the
model's own bremsstrahlung contributions by those of Low's theorem,
the results approach each other asymptotically towards the large-$s$
endpoint.  Without this substitution, the differential branching
ratios from HH$\chi$PT in this region would be about one order of
magnitude larger. For lower $s$, the impact of the resonances becomes
visible.

In the soft photon approximation the photon couples directly to the
mesons. Therefore, there is no such contribution for the $D \to \pi^0
\overline{K}^0 \gamma$ decay. Its distribution is dominated by the
$\omega$ resonance which has a significant branching ratio to $\pi^0
\gamma$; this is manifest in the Dalitz plot in
Fig.~\ref{fig:SM_d2BR}.

Apart from the $K^*, \rho$, and $\phi$ peaks, the shapes of the
differential branching ratios differ significantly between QCDF and
HH$\chi$PT, due to the $t$ and $u$-channel resonance contributions in
the latter. This is shown in the Dalitz plot in
Fig.~\ref{fig:SM_d2BR}.

In Table~\ref{tbl:branching_ratios} we give the SM branching ratios
for the four decay modes. We employ phase space cuts $s \leq
1.5~\GeV^2$, the region of applicability of QCDF, or
$E_\gamma\geq0.1~\GeV$, corresponding to $s \leq 3.1~\GeV^2$, to avoid
the soft photon pole. Here, $E_\gamma=(m_D^2-s)/(2 m_D)$ is the photon
energy in the $D$ meson's rest frame. Applying the same cuts in both
cases, the HH$\chi$PT branching ratios are generally larger than the
QCDF ones, except for the $D \to K^+ K^- \gamma$ mode, where they are
of comparable size.
 
We recall that SM branching ratios within leading order QCDF are
proportional to $(1/\lambda_D)^2$. Since $\lambda_D$ is of the order
of $\Lambda_{\rm QCD}$ and we employ a rather low value $\lambda_D =
0.1\, \GeV$~\cite{deBoer:2017que}, the values in
Table~\ref{tbl:branching_ratios} should be regarded as maximal
branching ratios. The large uncertainty of these values arises from
the residual scale dependence of the Wilson coefficient $\bar
C$~(\ref{eq:Cbar}).  A measurement of the branching ratios of the
SM-like modes $D \to K \pi \gamma$ thus provides an experimentally
extracted value of $\bar C/\lambda_D$.  Color-allowed modes feature
Wilson coefficients with significantly smaller scale uncertainty, and
allow for a cleaner, direct probe of
$\lambda_D$~\cite{deBoer:2017que}. While $\lambda_D$ is poorly known,
it effectively drives the annihilation with initial state radiation
and experimental constraints are informative even in the presence of
sizable systematic uncertainties inherent to QCDF in charm.

\begingroup
\renewcommand{\arraystretch}{1.4}
\begin{table}
  \centering
  \begin{tabular}{l|c|c|c|c}
      & $D \to \pi^0 \overline{K}^0 \gamma$ & $ D \to \pi^+ K^- \gamma$ & $D \to \pi^+ \pi^- \gamma$ & $D \to K^+ K^- \gamma$ \\
    \hline
    $\text{QCDF}\big|^\text{SM}_{s \leq 1.5~\GeV^2} $ & $(0.04 - 6.36)\cdot 10^{-5}$ & $(0.01 - 1.28)\cdot 10^{-4}$ & $(0.04 - 5.16)\cdot 10^{-6}$ & $(0.05 - 9.92)\cdot 10^{-6}$ \\
    $\text{HH$\chi$PT}\big|^\text{SM}_{s \leq 1.5~\GeV^2} $ & $(0.9 - 2.2)\cdot 10^{-3}$ & $(7.2 - 9.2)\cdot 10^{-5}$ & $(6.2 - 7.1)\cdot 10^{-6}$ & $(1.1 - 1.6)\cdot 10^{-6}$ \\
    $\text{HH$\chi$PT}\big|^\text{SM}_{E_\gamma \geq 0.1~\GeV} $ & $(2.1 - 5.0)\cdot 10^{-3}$ & $(6.7 - 7.2)\cdot 10^{-4}$ & $(3.9 - 4.1)\cdot 10^{-5}$ & $(3.2 - 3.5)\cdot 10^{-5}$ \\
    $\text{QCDF}\big|^\text{BSM}_{s \leq 1.5~\GeV^2} $ & - & - & $(0.6 - 1.7)\cdot 10^{-5}$ & $(0.1 - 10.5)\cdot 10^{-6}$ \\
    $\text{HH$\chi$PT}\big|^\text{BSM}_{s \leq 1.5~\GeV^2} $ & - & - & $(0.9 - 1.7)\cdot 10^{-5}$ & $(0.9 - 1.7)\cdot 10^{-6}$ \\
    $\text{HH$\chi$PT}\big|^\text{BSM}_{E_\gamma \geq 0.1~\GeV} $ & - & - & $(4.3 - 5.3)\cdot 10^{-5}$ & $(3.2 - 3.6)\cdot 10^{-5}$
  \end{tabular}
  \caption{SM and BSM branching ratios for $D \to \pi^0 \overline{K}^0
    \gamma$ , $ D \to \pi^+ K^- \gamma$, $D \to \pi^+ \pi^- \gamma$
    and $D \to K^+ K^- \gamma$.  QCDF is applicable for $s \lesssim
    1.5\, \GeV^2$; to enable sensible comparison we also provide
    HH$\chi$PT branching ratios with this cut.  Also given are
    HH$\chi$PT predictions for $E_\gamma \geq 0.1\, \GeV$, see text
    for details. The QCDF branching ratios are obtained for
    $\lambda_D=0.1\, \GeV$. The SM predictions are $\propto (0.1\,
    \GeV/\lambda_D)^2$.}
  \label{tbl:branching_ratios}
\end{table}
\endgroup

\subsection{Forward-Backward Asymmetry\label{sec:AFB}}

\begin{figure}
  \centering
  \includegraphics[width=0.9\linewidth]{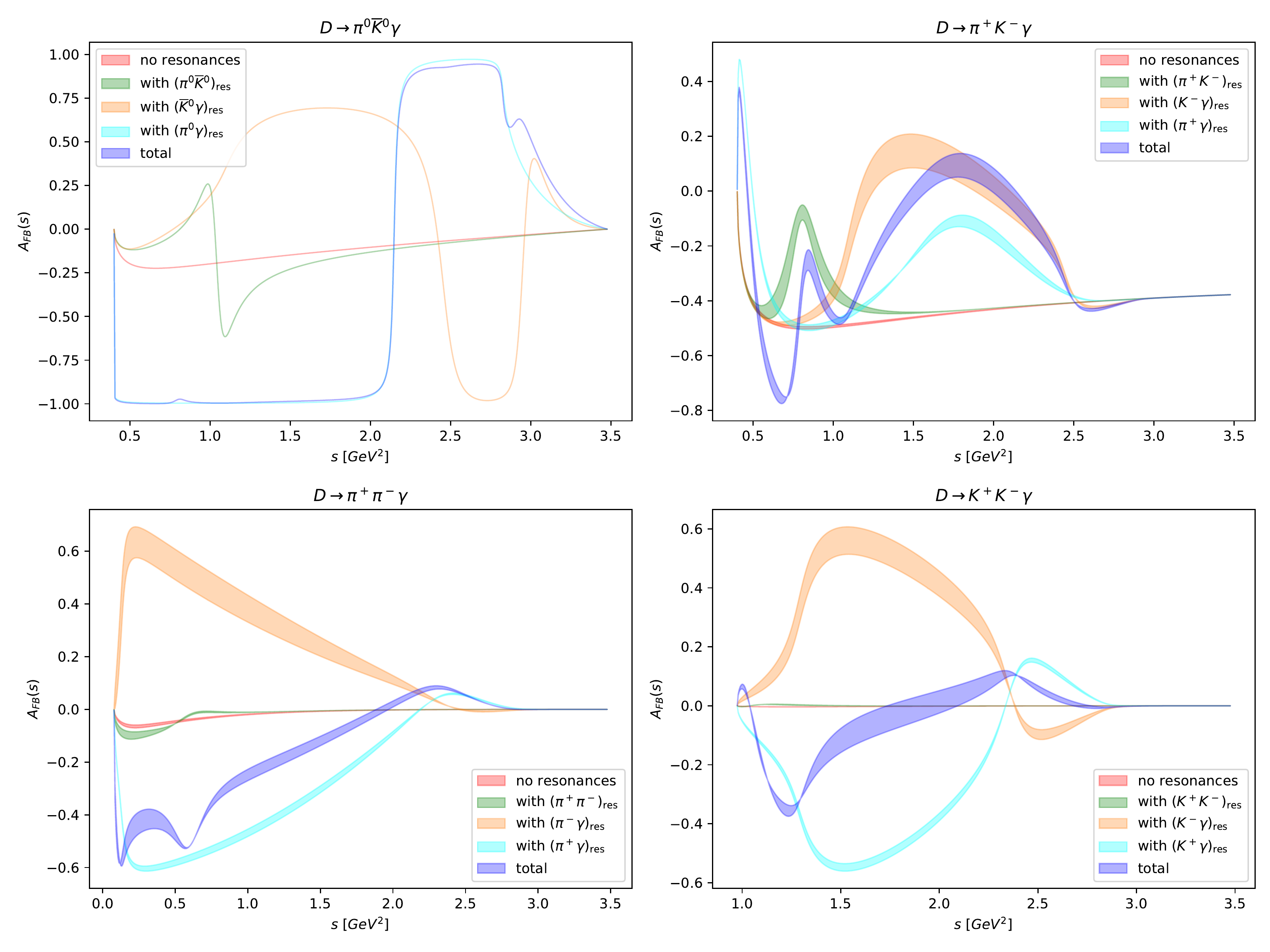}
  \caption{The forward-backward asymmetry $A_{\rm FB}(s) $
    (\ref{eq:AFB}) as a function of $s$. The red bands contain only
    non-resonant contributions. The green, orange and light blue bands
    contain additional contributions of a specific resonance
    channel. The dark blue bands are the complete forward-backward
    asymmetries according to HH$\chi$PT.  To leading order QCDF
    $A_{\rm FB}(s)=0 $.}
  \label{fig:A_FB_SM}
\end{figure}

Angular observables are also suitable for testing QCD models. We
define the forward-backward asymmetry
\begin{align} \label{eq:AFB}
  \begin{split}
    &A_{\rm FB}(s) = \frac{\int_{t_{\rm min}}^{t_{\rm 0}} dt \frac{\dd{}^2\Gamma}{\dd{}s \dd{}t} -\int_{t_{\rm 0}}^{t_{\rm max}} dt  \frac{\dd{}^2\Gamma}{\dd{}s \dd{}t}}{\int_{t_{\rm min}}^{t_{\rm 0}} dt \frac{\dd{}^2\Gamma}{\dd{}s \dd{}t} + \int_{t_{\rm 0}}^{t_{\rm max}} dt  \frac{\dd{}^2\Gamma}{\dd{}s \dd{}t}} \, ,\\
  &t_{\rm 0} = \frac{1}{2s}\left(-s^2 + s(m_D^2 + m_1^2 + m_2^2) + m_D^2(m_2^2 - m_1^2)\right)\, ,
  \end{split}
\end{align}
where the first (second) term in the numerator corresponds to $0 \leq
\cos(\theta_{2\gamma}) \leq 1$ $(-1 \leq \cos(\theta_{2\gamma}) \leq
0)$. Here, $\theta_{2\gamma}$ is the angle between $P_2$ and the
photon in the $P_1 - P_2$ center-of-mass frame. In
Fig.~\ref{fig:A_FB_SM} we show the SM forward-backward asymmetry based
on HH$\chi$PT. In all decay modes $A_{\rm FB}(s)$ is dominated by
intermediate vector resonances. To illustrate this, the
forward-backward asymmetries are also shown without or only with
individual resonance contributions. The $(P_1P_2)_\text{res}$
resonances contribute to $A_{\rm FB}$ only via interference terms,
since the corresponding form factors depend only on $s$. For $D \to
\pi^+ \pi^- \gamma$ and $D \to K^+ K^- \gamma$ the diagrams of the
neutral current operator, which contain $(P_1\gamma)_\text{res}$ and
$(P_2\gamma)_\text{res}$ resonances, give the same contribution to the
amplitude in the forward and backward region of the phase space. For
$P_1 \neq \overline{P}_2$ this symmetry does not exist. In case of the
charged current operator, these resonances contribute in different
ways to the forward and backward region due to the asymmetric
factorization of the diagrams $B_3$ (\ref{eq:BpiK}), (\ref{eq:Bpipi}),
(\ref{eq:BKK}).  This effect is primarily responsible for the shape of
$A_{\rm FB}$ in $D \to \pi^+ \pi^- \gamma$ and $D \to K^+ K^- \gamma$
decays. $A_{\rm FB}(D \to \pi^0 \overline{K}^0 \gamma)$ is, like the
differential branching ratio shown in Fig.~\ref{fig:SM_d2BR},
dominated by the $\omega$ resonance.

Since the WA form factors are only dependent on $s$, the SM
forward-backward asymmetry vanishes to leading order QCDF. Therefore,
we add contributions from $t$ and $u$-channel resonances using a
phenomenological approach. To this end, we combine $D \to VP$
amplitudes with the effective $VP\gamma$ coupling from equation
\eqref{eq:L_VPgamma}.
\begin{figure}
  \centering
  \includegraphics[width=0.8\linewidth]{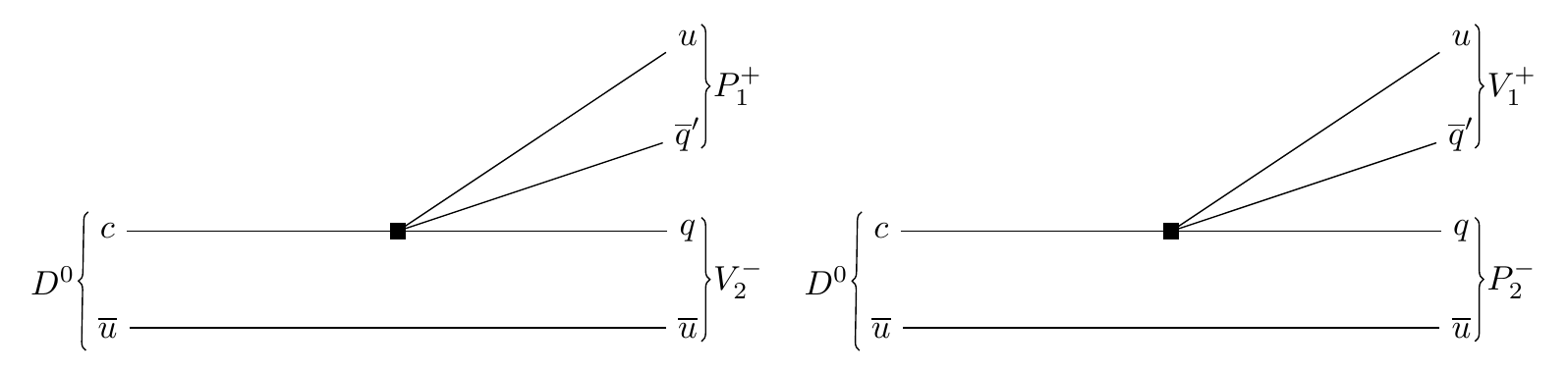}
  \caption{The dominant diagrams to $D\to VP(\to P P \gamma)$ in the
    $\lambda_\text{QCD}/m_c$ and $\alpha_s$ expansion. At order
    $\alpha_s^0$ QCDF reproduces the naive
    factorization~\cite{Beneke:2000ry}. The diagrams are shown for
    charged final state mesons. For the final state with uncharged
    mesons the $u$ and $q$ quark have to be exchanged.}
  \label{fig:D to PV}
\end{figure}
We obtain
\begin{align}
  \begin{split}
    \A^\text{pheno}_+ = \frac{G_F e}{\sqrt{2}}V_{cq}^* V_{uq'} \left(C_2 - \frac{1}{6}C_1\right)\left(\frac{2m_V f_P g_{VP\gamma}A_0^{DV}(m_1^2)}{(p_2+k)^2 - m_V^2 + \im m_V \Gamma_V}+ \frac{2m_V f_V g_{VP\gamma}F_1^{DP}((p_1+k)^2)}{(p_1+k)^2 - m_V^2 + \im m_V \Gamma_V}\right) \, , \label{eq:A_pheno}
  \end{split}
\end{align}
where the first (second) term in \eqref{eq:A_pheno} corresponds to the
left (right) diagram in Fig.~\ref{fig:D to PV}. The amplitude for the
final state $\pi^0 \overline{K}^0 \gamma$ can be obtained from
Eq~\eqref{eq:A_pheno} by substituting $C_2-1/6C_1 \rightarrow C_2/2$,
$m_1 \to m_2$, and $p_1 \leftrightarrow p_2$, and multiplying by the
factor $-1/\sqrt{2}$. The $D\to P$ and $D\to V$ transition form
factors are taken from Ref.~\cite{Verma:2011yw}. As expected,
resulting distributions based on~\eqref{eq:A_pheno} exhibit the same
main resonance features as the ones in HHchiPT, and are therefore not
shown.

\subsection{The U-spin link \label{sec:U}}

We further investigate the U-spin link between the SM-dominated mode
$D \to K^- \pi^+ \gamma$ and the BSM-probes $D \to \pi^+ \pi^- \gamma$
and $D \to K^+ K^- \gamma$.  In practise, a measurement of ${\cal{
    B}}(D \to K^- \pi^+ \gamma)$ can provide a data-driven SM
prediction for the branching ratios of the FCNC decays.  The method is
phenomenological and serves, in the case of branching ratios, as an
order-of-magnitude estimate.  The U-spin approximation is expected to
yield better results in ratios of observables (which arise already at
lowest order in the U-spin limit), such that overall systematics drops
out.  Useful applications have been made for polarization asymmetries
in $D \to V \gamma$ decays \cite{deBoer:2018zhz}.  However, three-body
radiative decays are considerably more complicated due to the
intermediate resonances, and we do not pursue the U-spin link for the
forward-backward or CP asymmetries.

\begin{figure}
  \centering
  \includegraphics[width=1\linewidth]{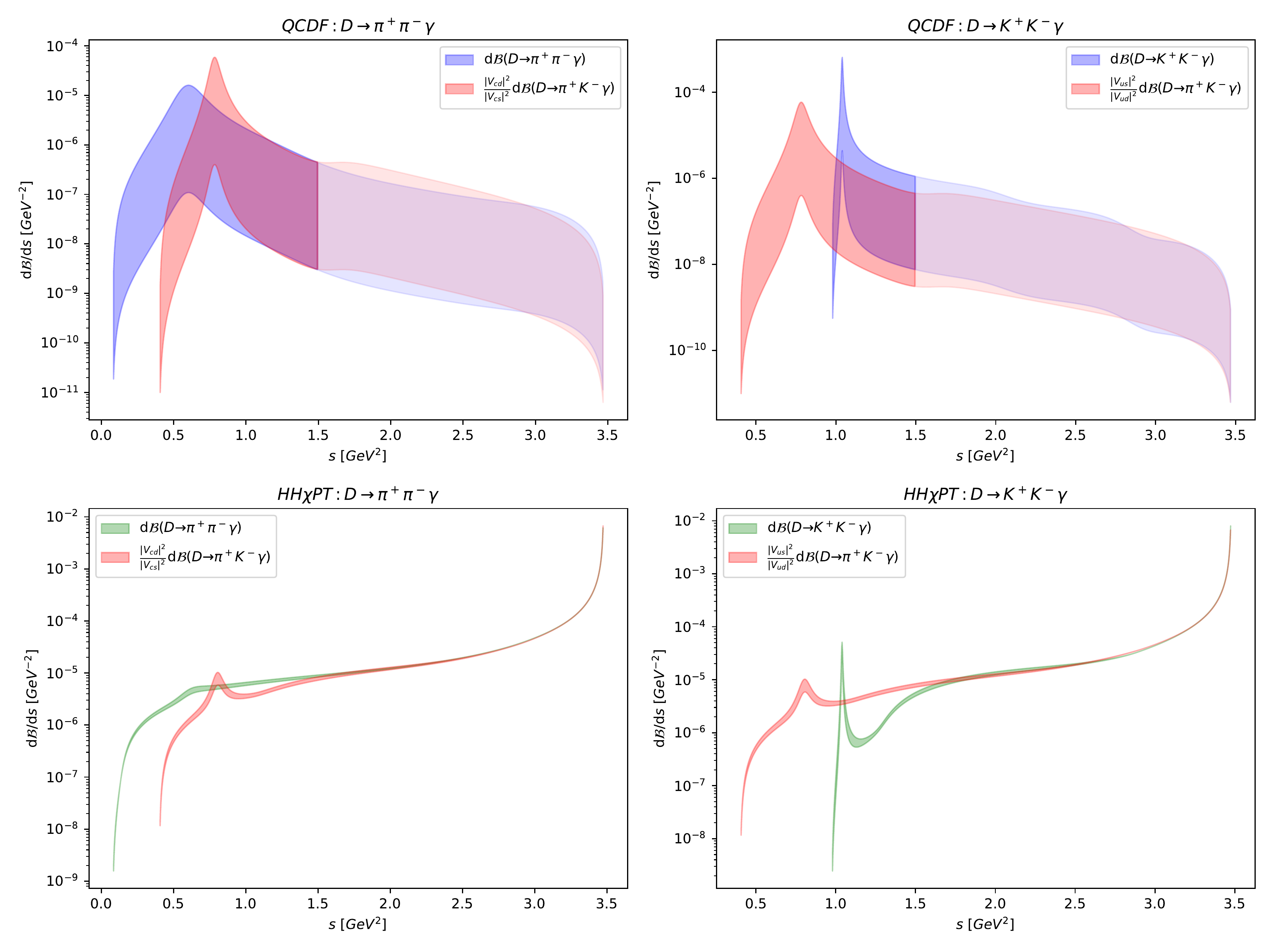}
  \caption{The SM predictions for the differential branching ratios of the decays  $D \to \pi^+ \pi^- \gamma$ (left) and $D \to K^+ K^- \gamma$ (right)
  from a direct QCDF computation (blue bands in upper plots), HH$\chi$PT computations (green bands in lower plots) and from the $D \to K^- \pi^+ \gamma$ distribution multiplied by $|V_{cd}/V_{cs}|^2$ and $|V_{us}/V_{ud}|^2$, respectively (red bands).
  The prediction for the SM-like mode $D \to K^- \pi^+ \gamma$ in this figure is from the respective models but could be taken from data.}
  \label{fig:Uspin}
\end{figure}

A comparison between $|V_{us}|^2/|V_{ud}|^2 d{\cal{ B}} (D \to K^-
\pi^+ \gamma)/ds$ with $d{\cal{ B}} (D \to K^+ K^- \gamma)/ds$ and
$|V_{cd}|^2/|V_{cs}|^2 d{\cal{ B}}(D \to K^- \pi^+ \gamma)/ds$ with
$d{\cal{ B}} (D \to \pi^+ \pi^- \gamma)/ds$ is shown in
Fig.~\ref{fig:Uspin}.  For $s\gtrsim1.5~\GeV$ the predictions of the
direct calculations and the U-spin relations are in good
agreement. This holds for both the extrapolations of QCDF and the
HH$\chi$PT predictions. In the second case this is due to the
dominance of the bremsstrahlung contributions and the U-spin relations
of the $D\to P_1 P_2$ amplitudes. For $s \lesssim 1.5~\GeV$, there are
large deviations due to the differences in phase space boundaries and
the different intermediate resonances.
At the level of integrated SM branching ratios we find  
\begin{align} 
&\frac{ {\cal{B}} - {\cal{B}}(\text{U-spin link})}{{\cal{B}}}  \big \vert^\text{QCDF}_{s \leq 1.5~\GeV^2}  \sim -0.33 \,  (0.3) \, , \label{eq:Ubreak1} \\
&\frac{ {\cal{B}} - {\cal{B}}(\text{U-spin link})}{{\cal{B}}} \big  \vert^\text{HH$\chi$PT}_{s \leq 1.5~\GeV^2}   \sim 0.35  \,  (-2.3) \, , \label{eq:Ubreak2} \\
&\frac{ {\cal{B}} - {\cal{B}}(\text{U-spin link})}{{\cal{B}}} \big  \vert^\text{HH$\chi$PT}_{E_\gamma \geq 0.1~\GeV}   \sim 0.07  \,  (-0.11) \, ,\label{eq:Ubreak3} 
\end{align}
for the $D \to \pi^+ \pi^- (K^+ K^-)\gamma$
modes. Eqs.~\eqref{eq:Ubreak1}-\eqref{eq:Ubreak3} underline the main
features of Fig~\ref{fig:Uspin}: as a result of the dominance of
bremsstrahlung photons from Low’s theorem the
corrections~\eqref{eq:Ubreak3} are small; the proximity of the $\phi$
to the phase space boundary in $D \to KK \gamma$ makes the U-spin
limit in~\eqref{eq:Ubreak2} poor. In the other cases the U-spin
symmetry performs as expected, within $\sim 30$\%.

\section{BSM analysis \label{sec:BSM}}

BSM physics can significantly increase the Wilson coefficients
contributing to $c \to u \gamma$ transitions.  Examples are
supersymmetric models with flavor mixing and chirally enhanced gluino
loops, or leptoquarks, see Ref.~\cite{deBoer:2017que} for details. In
the following we work out BSM spectra and phenomenology in a
model-independent way. Experimental data obtained from $D \rightarrow
\rho^0 \gamma$ decays provide model-independent constraints
\cite{deBoer:2018buv, Abdesselam:2016yvr}
\begin{align} \label{eq:max}
  |C_7|, |C_7^\prime| \lesssim 0.3\, .
\end{align}
These values are in agreement with recent studies of $D \to \pi l l$
decays~\cite{Bause:2019vpr}. In Section~\ref{sec:c8} we discuss the
implications of CP asymmetries in hadronic charm decays that can lead
to constraints on the imaginary parts of the dipole operators.

The $D \to P_1 P_2$ matrix elements of the tensor currents can be parameterized as
\begin{align}
  \braket{P_1(p_1)P_2(p_2)| \ubar \sigma^{\mu \nu} k_\mu (1\pm \gamma_5) c| D(P)} = m_D \left[a^\prime p_1^\nu + b^\prime p_2^\nu + c^\prime P^\mu \mp 2 i  h^\prime \epsilon^{\nu\alpha\beta\gamma} p_{1\alpha} p_{2\beta} k_\gamma\right]\, . \label{eq:Tensor_Formfaktoren}
\end{align}
with the form factors $a^\prime, b^\prime, c^\prime, h^\prime$ given in  App.~\ref{app:HQCHPT form factors}. 
The form factors depend on $s$ and $t$ and satisfy
\begin{align}
a' p_1 \cdot k + b' p_2 \cdot k + c' P \cdot k=0 \, . 
\end{align}
  The BSM amplitudes are then obtained as
\begin{align} 
  \begin{split} 
    &\A_-^{\rm BSM} = i \frac{G_F e}{\sqrt{2}} \frac{m_c }{4\pi^2 }(C_7 + C_7^\prime)\frac{(b^\prime - a^\prime)}{v\cdot k} \, ,\\
    &\A_+^{\rm BSM} = \frac{G_F e}{\sqrt{2}} \frac{m_c m_D}{2\pi^2 } (C_7 - C_7^\prime) h^\prime\, . \label{eq:BSM-Amplituden}
  \end{split}
\end{align}

\begin{figure}
  \centering
  \includegraphics[width=\linewidth]{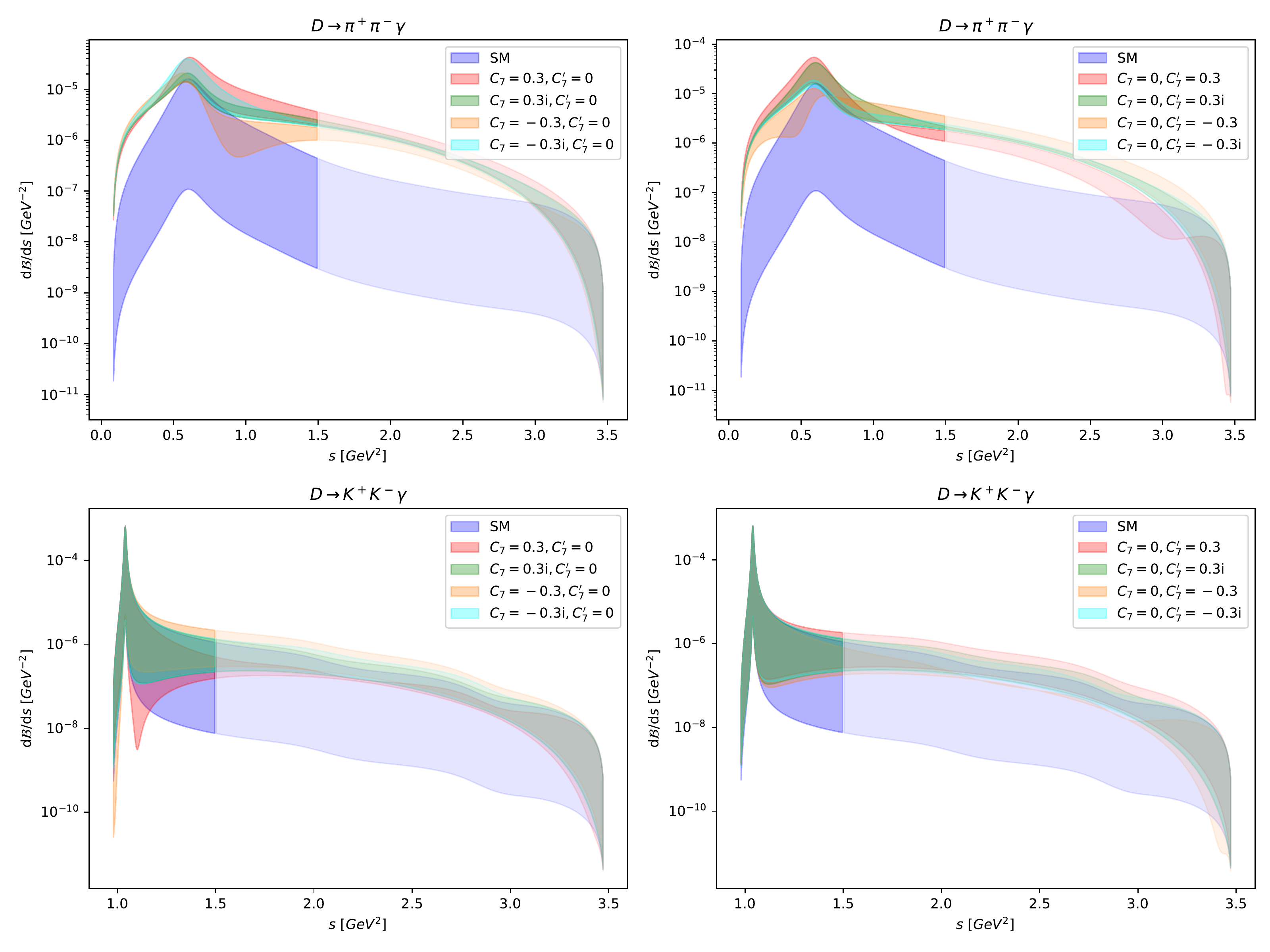}
  \caption{Comparison of QCDF-based SM predictions of differential
    branching ratios for $D \to \pi^+ \pi^- \gamma$ (upper plots) and
    $D \to K^+ K^- \gamma$ (lower plots) within different BSM
    scenarios. One BSM coefficient is set to zero while the other one
    exhausts the limit (\ref{eq:max}) with CP-phase $0, \pm \pi/2,
    \pi$.}
  \label{fig:d1BR_QCDF_BSM}
\end{figure}

\begin{figure}
  \centering
  \includegraphics[width=\linewidth]{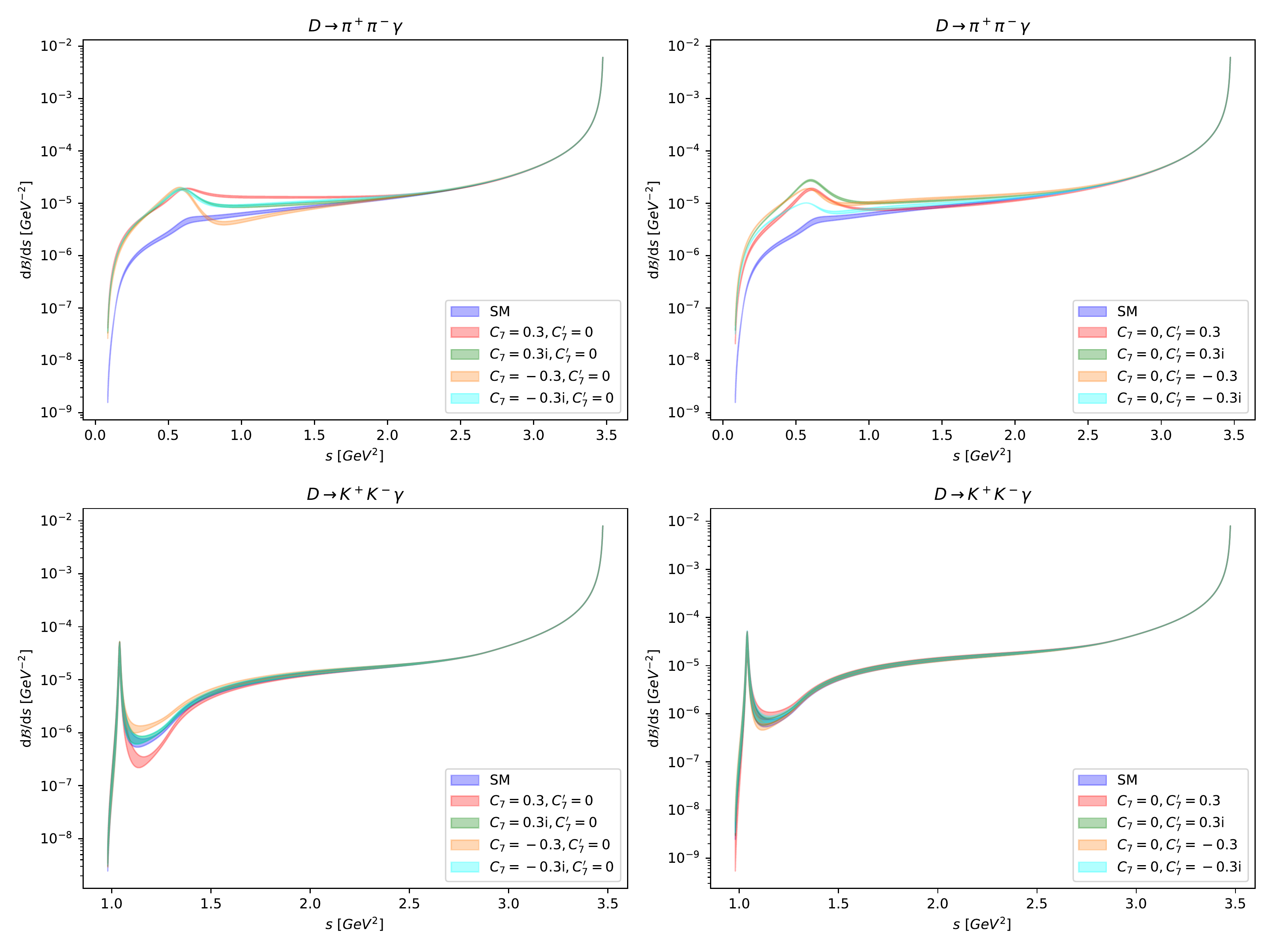}
  \caption{As in Fig.~\ref{fig:d1BR_QCDF_BSM} but for  HH$\chi$PT.}
   \label{fig:d1BR_HQCHIPT_BSM}
\end{figure}

In Figs.~\ref{fig:d1BR_QCDF_BSM} and~\ref{fig:d1BR_HQCHIPT_BSM} we
show differential branching ratios for the FCNC modes based on QCDF
and HH$\chi$PT, respectively, both in the SM (blue) and in different
BSM scenarios. One of the BSM coefficients, $C_7$ or $C_7^\prime$, is
set to zero while the other one is taken to saturate the limit
(\ref{eq:max}) with CP-phases $0, \pm \pi/2, \pi$. The same
conclusions are drawn for both QCD approaches: the $D\to K^+ K^-
\gamma$ branching ratio is insensitive to NP in the dipole
operators. In particular, the benchmarks for $O_7^\prime$ and the SM
prediction are almost identical. For $O_7$ small deviations occur
directly beyond the $\phi$ peak. On the other hand, BSM contributions
can increase the differential branching ratio of $D \to \pi^+ \pi^-
\gamma$ by up to one order of magnitude around the $\rho$ peak.
However, due to the intrinsic uncertainties from the Breit-Wigner
contributions around the resonance peaks it is difficult to actually
claim sensitivity to NP. This is frequently the case in $D$ physics
for simple observables such as branching ratios. The NP sensitivity is
higher in observables involving ratios, such as CP asymmetries,
discussed in the next section. 

\begin{figure}
  \centering
  \includegraphics[width=0.8\linewidth]{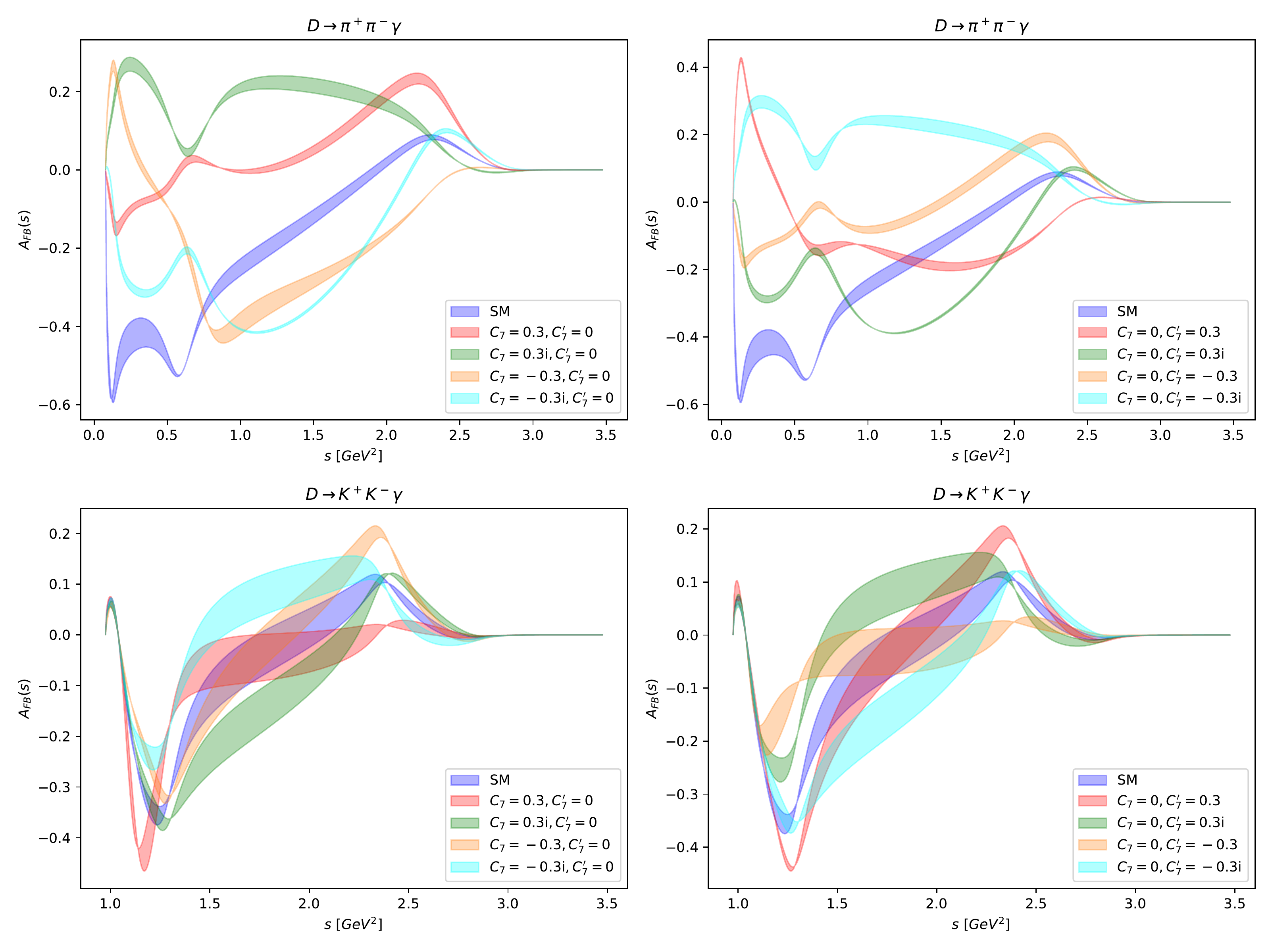}
  \caption{The forward-backward asymmetry in the SM (blue band) and
    beyond for the decays $D \to \pi^+ \pi^- \gamma$ and $D \to K^+
    K^- \gamma$ as a function of $s$, based on the HH$\chi$PT form
    factors.}
  \label{fig_FB_asymmetrie_HHCHIPT}
\end{figure}

\begin{figure}
  \centering
  \includegraphics[width=0.8\linewidth]{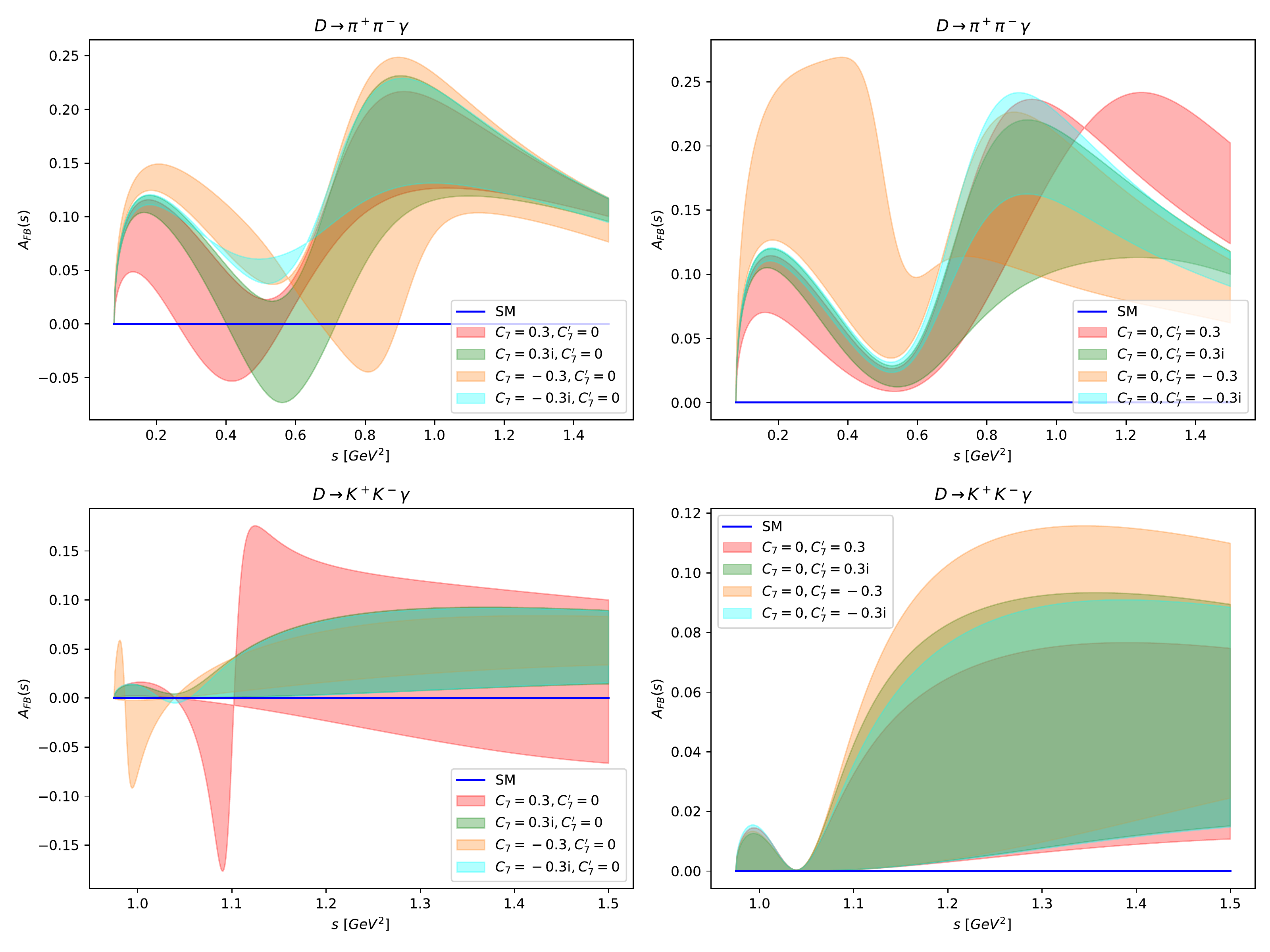}
  \caption{As in Fig.~\ref{fig_FB_asymmetrie_HHCHIPT} but within QCDF
    (\ref{eq:BSM-Amplituden}). }
    \label{fig_FB_asymmetrie_QCDF}
\end{figure}

The NP impact on $A_{\rm FB}$ is sizable, see
Fig.~\ref{fig_FB_asymmetrie_HHCHIPT} for the HH$\chi$PT predictions.
However, due to the complicated interplay of $s$-, $t$- and
$u$-channel resonances further study in SM-like $D \to K \pi \gamma$
decays is suggested to understand the decay dynamics before drawing
firm conclusions within NP.
Since the form factors depend on $s$ and $t$, the pure BSM
contributions (\ref{eq:BSM-Amplituden}) induce a forward-backward
asymmetry within QCDF, whereas it vanishes in the SM (see
Fig.~\ref{fig_FB_asymmetrie_QCDF}).

\section{CP Violation \label{sec:CP}}

\begin{figure}
  \centering
  \includegraphics[width=\linewidth]{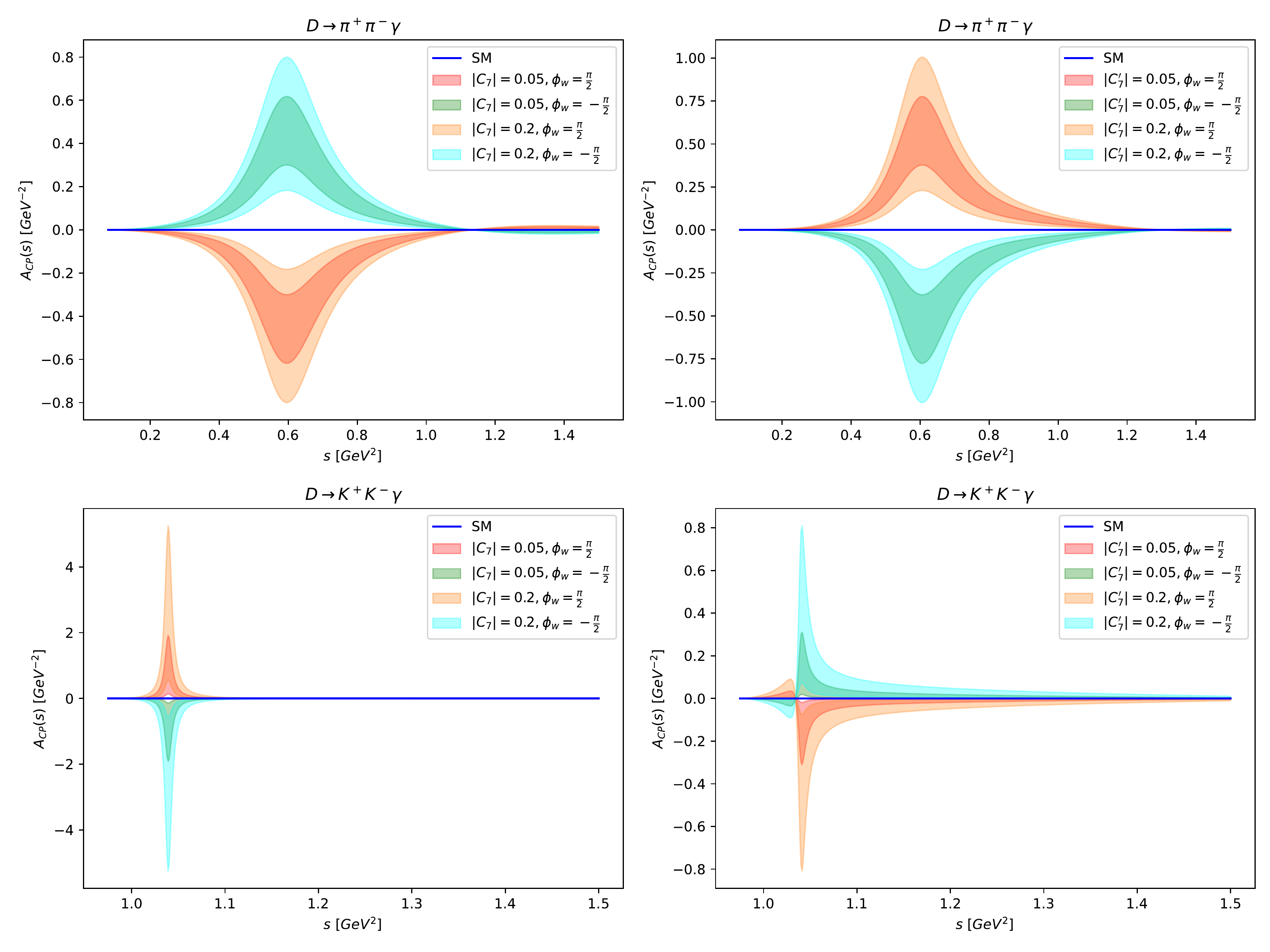}
  \caption{Predictions for the CP asymmetries in $D \to \pi^+ \pi^-
    \gamma$ and $D \to K^+ K^- \gamma$ as a function of $s$, within
    the SM and beyond (using~\eqref{eq:BSM-Amplituden}), based on
    QCDF.  For the BSM scenarios, we have set one coefficient
    $C_7^{(\prime)}$ to 0 and the other one to $0.05, 0.2$. The weak
    phase of $C_7^{(\prime)}$ is $\phi_w=\pm \pi/2$. We performed a
    cut $s \leq 1.5\, \GeV^2$ to remain within the region where QCDF
    applies.}
  \label{fig:CP_asymmetrie_QCDF}
\end{figure}

\begin{figure}
  \centering
  \includegraphics[width=\linewidth]{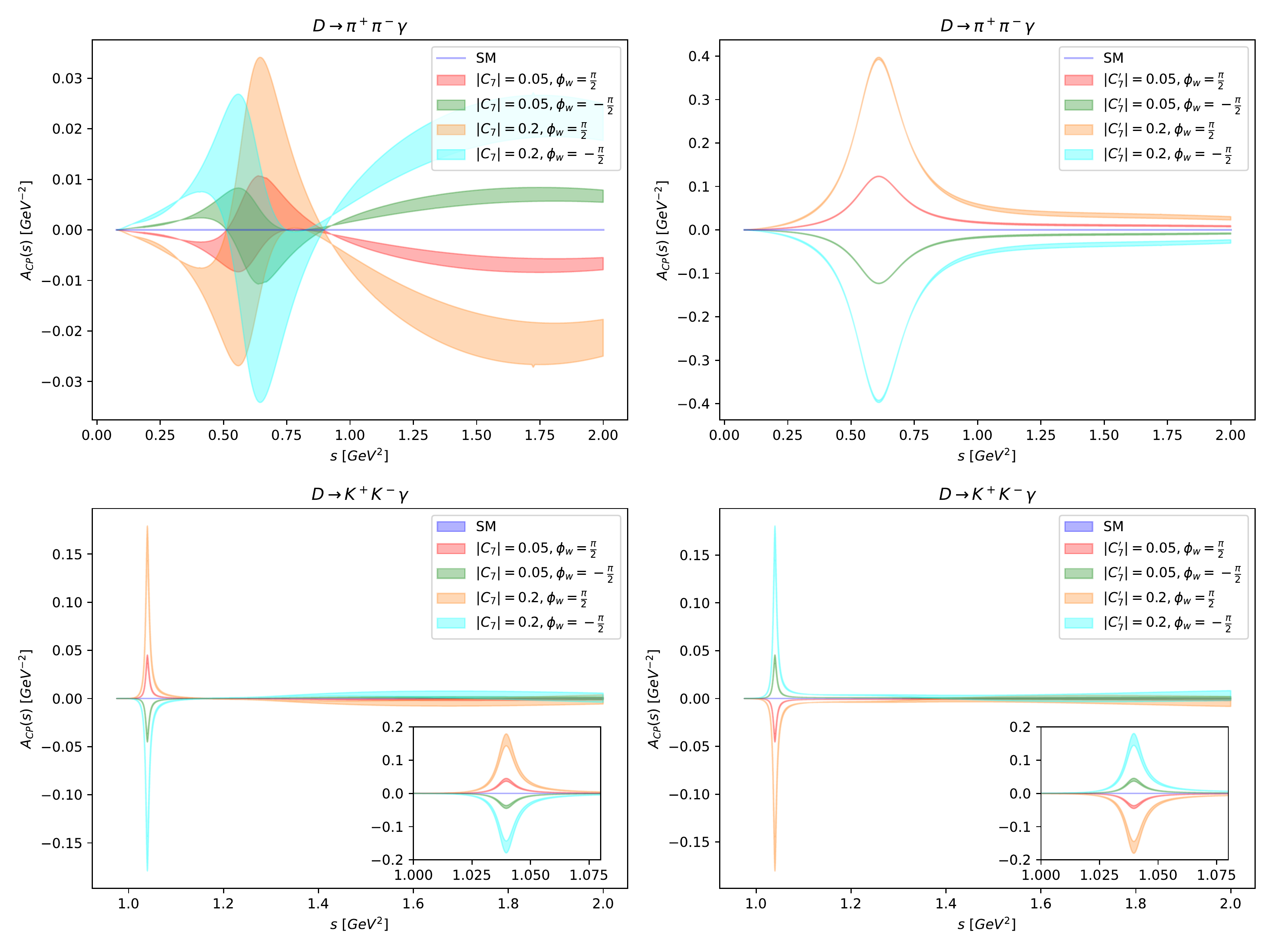}
  \caption{ As in Fig.~\ref{fig:CP_asymmetrie_QCDF} but for HH$\chi$PT and with cut $s \leq 2\, \GeV^2$ to avoid large bremsstrahlung contributions in the normalization.}
  \label{fig:CP_asymmetrie_HHchiPT}
\end{figure}

Another observable that offers the possibility to test for BSM physics
is the single- or double-differential CP asymmetry. It is defined,
respectively, by
\begin{align}
 A_{\rm CP}(s) =\int \dd{} t    A_{\rm CP}(s,t) \, , ~~~
  A_{\rm CP}(s,t) =\frac{1}{\Gamma + \overline{\Gamma}}\left(\frac{\dd{}^2\Gamma}{\dd{}s\dd{}t} - \frac{\dd{}^2\overline{\Gamma}}{\dd{}s\dd{}t}\right) \, . 
\end{align}
Here, $\overline{\Gamma}$ refers to the decay rate of the
CP-conjugated mode.
Within the SM, $D \to K^+ K^- \gamma$ is the only decay that contains
contributions with different weak phases and thus the only decay mode
with a nonvanishing CP asymmetry. A maximum of $A^{\text{SM}}_{\rm
  CP}(s) \lesssim 1.4 \cdot 10^{-4}$ located around the $\phi$ peak is
predicted by QCDF. Since the $\phi$ is a narrow resonance, the CP
asymmetry decreases rapidly with increasing $s$. BSM contributions can
contain further strong and weak phases and thus significantly increase
the CP asymmetry. In Fig.~\ref{fig:CP_asymmetrie_QCDF} we show the
predictions for the CP asymmetries within the SM and for several
different BSM scenarios, based on QCDF. We assign a non-zero value to
one of the BSM coefficients and set the weak phase to $\phi_w=\pm
\pi/2$. The BSM CP asymmetries $A_{\rm CP}(s)$ can, in principle,
reach $\O(1)$ values. Constraints can arise from data on CP
asymmetries in hadronic decays; these are further discussed in
Section~\ref{sec:c8}. We emphasize that $A_{\rm CP}$ depends on cuts
used in the normalization $\Gamma+ \bar \Gamma$. In
Fig.~\ref{fig:CP_asymmetrie_QCDF} we include the contributions up to
$s=1.5~\GeV^2$.\\
\newpage
HH$\chi$PT predicts a SM CP asymmetry $A_{\rm CP}^\text{SM}(s)
\lesssim 0.7\cdot 10^{-4}$ for the $D \to K^+ K^- \gamma$ decay. In
Fig.~\ref{fig:CP_asymmetrie_HHchiPT} we show the same BSM benchmarks
as before, employing HH$\chi$PT. We performed a cut $s\leq 2\, \GeV$
to avoid large bremsstrahlung effects in the normalization, which
would artificially suppress $A_{\rm CP}$. Still, the CP asymmetries
obtained using HH$\chi$PT are smaller than those using QCDF, since a
larger part of the phase space is included in the normalization. 

\begin{figure}
  \centering
  \includegraphics[width=\linewidth]{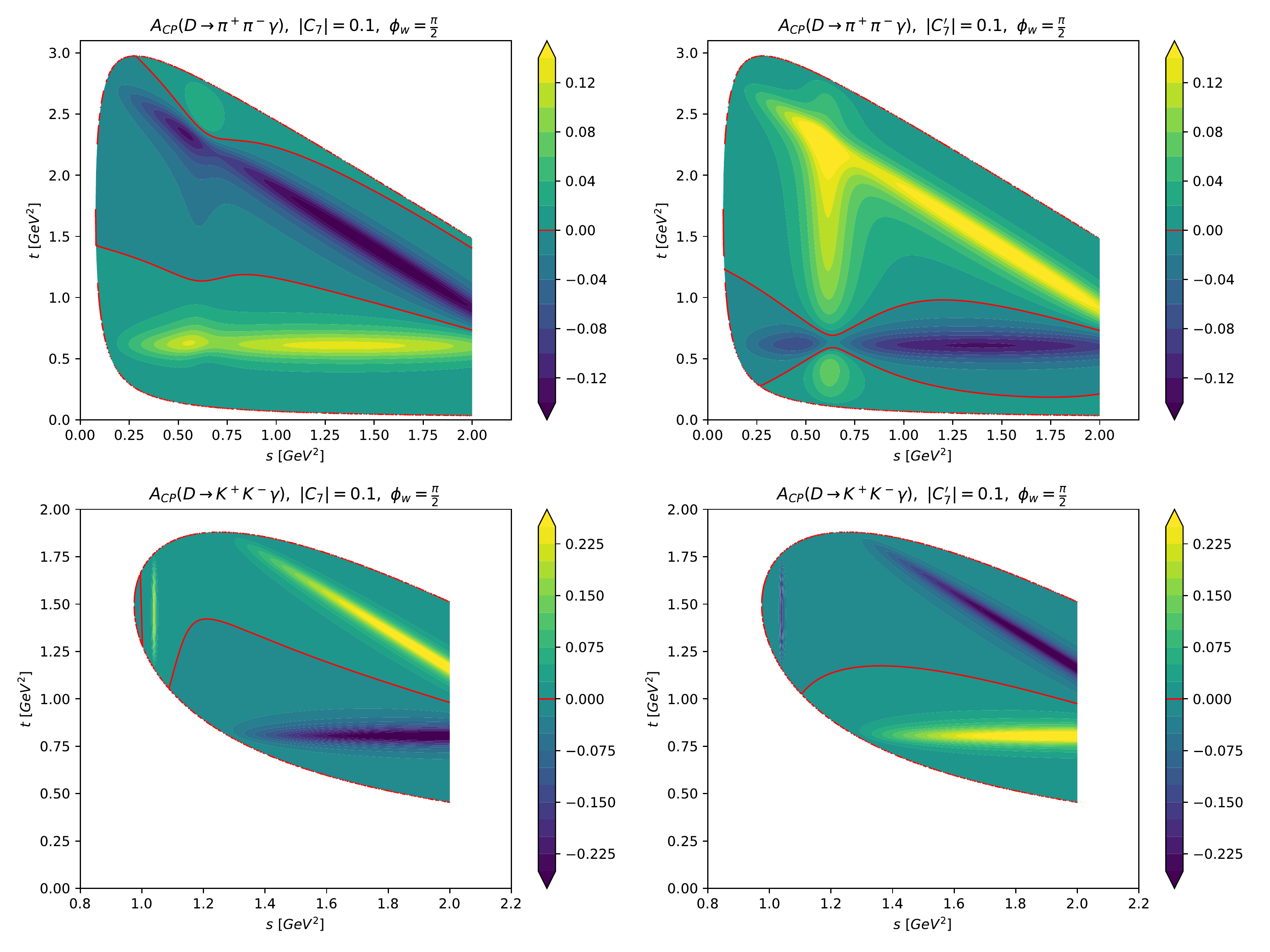}
  \caption{Dalitz plot of $A_{\rm CP}(s,t)$ for $D \to \pi^+ \pi^-
    \gamma$ (upper plots) and $D \to K^+ K^- \gamma$ decays (lower
    plots) based on HH$\chi$PT.  We have set one BSM coefficient,
    $C_7$ or $C_7^{\prime}$, to 0 and the other one to $0.1$, with
    weak phase $\phi_w= \pi/2$. We employed a cut $s \leq 2\, \GeV^2$
    to avoid large bremsstrahlung contributions in the normalization.}
  \label{fig:d2A_CP}
\end{figure}

For $D \to \pi^+ \pi^- \gamma$, the contributions of $\A_-$ and $\A_+$
to the CP asymmetries are of roughly the same size. Therefore, the
relative signs of the dipole Wilson coefficients
in~\eqref{eq:BSM-Amplituden} results in a constructive increase (for
$C_7^\prime$) and a cancellation (for $C_7$), respectively, of the CP
asymmetry. For the $D \to K^+ K^- \gamma$ mode, the $\phi$ resonance
contributes only to $A_+$. Therefore, in this case the CP asymmetry is
dominated by the parity-even amplitude.
In order to get additional strong phases and thus an increase of the
CP asymmetry, one could consider further heavy vector resonances such
as the $\phi(1680)$. Intermediate scalar particles like $f_0(1710)$
\cite{Soni:2019xko} would also add additional strong phases. We remark
that $A_{\rm CP}$ can change its sign in dependence of $s$; therefore,
binning is required to avoid cancellations. $A_{\rm CP}$ is very small
beyond the $(P_1P_2)_\text{res}$ peak due to the cancellation of the
$(P_1 \gamma)_\text{res}$ and $(P_2 \gamma)_\text{res}$ contributions
upon integration over $t$. To avoid this cancellation one could use
the $s$- and $t$-dependent CP asymmetry $A_{\rm CP}(s,t)$ as shown in
Fig.~\ref{fig:d2A_CP}. Note that part of the resonance contribution to
the asymmetry is removed by the bremsstrahlung cut.

\begin{figure}
  \centering
  \includegraphics[width=0.54\linewidth]{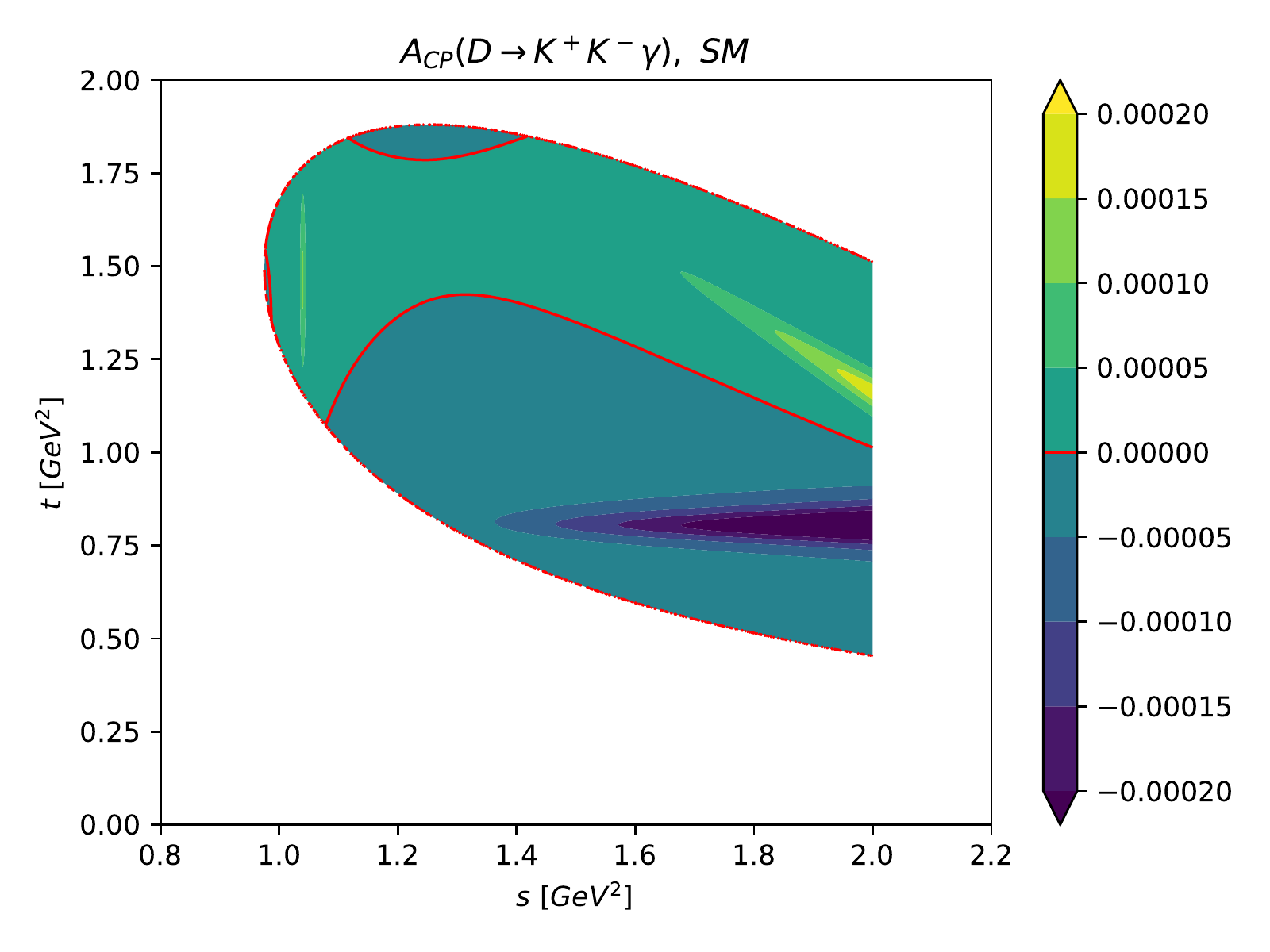}
  \caption{Dalitz plot of $A_{\rm CP}^{\rm SM}(s,t)$ for $D \to K^+
    K^- \gamma$ decays based on HH$\chi$PT for $s \leq 2\, \GeV^2$.}
  \label{fig:st-SM}
\end{figure}

\subsection{CP phases and $\Delta A_{\rm CP}$ \label{sec:c8}}

We briefly discuss the impact of the chromomagnetic dipole operators
$O_8^{(\prime)}$ on radiative charm decays, where
\begin{equation}
       O_8 = \frac{g_s m_c}{16\pi^2}  \left(\ubar_L \sigma^{\mu \nu} G_{\mu \nu} c_R\right)\,,
\qquad O_8^\prime = \frac{g_s m_c}{16\pi^2}  \left(\ubar_R \sigma^{\mu \nu} G_{\mu \nu}  c_L\right)\, ,
\end{equation}
and $G_{\mu \nu}$ denotes the chromomagnetic field strength tensor.
We do not consider contributions from $O_8^{(\prime)}$ to the matrix
element of $D \to PP \gamma$ decays, which is beyond the scope of this
work. The corresponding contributions for the $D \to V \gamma$ decays
have been worked out in Ref.~\cite{deBoer:2017que}.

The QCD renormalization-group evolution connects the electromagnetic
and the chromomagnetic dipole operators at different scales. To
leading order we find the following relation~\cite{deBoer:2017que},
\begin{align} \label{eq:rglink}
C_7^{(\prime)}(m_c) \simeq 0.4 \big( C_7^{(\prime)}(\Lambda) -  C_8^{(\prime)}(\Lambda) \big) \,, 
\qquad C_8^{(\prime)}(m_c) \simeq  0.4 C_8^{(\prime)}(\Lambda)   \,,
\end{align}
which is valid to roughly 20\% if $\Lambda$, the scale of NP, lies
within 1-10 TeV. It follows that CP asymmetries for radiative decays
are related to hadronic decays, a connection discussed in
\cite{Isidori:2012yx, Lyon:2012fk} in the context of $\Delta A_{\rm CP}=A_{\rm
  CP}(D\to K^+ K^-)-A_{\rm CP}(D \to \pi^+\pi^-)$.  The latter is
measured by LHCb, $\Delta A_{\rm CP}=-(15.4 \pm 2.9) \cdot
10^{-4}$~\cite{Aaij:2019kcg}, and implies $\Delta A_{\rm CP}^{\rm NP}
\sim {\rm Im }(C_8-C_8^\prime) \sin \delta \lesssim 2 \cdot 10^{-3}$
for NP from dipole operators, with a strong phase difference $\delta$
and Wilson coefficients evaluated at $\mu=m_c$. For $\sin \delta \sim
O(1)$, and $C_8$ only (or $C_8^\prime$ only), strong constraints on
the electromagnetic dipole operators follow from (\ref{eq:rglink}),
unless $C_7(\Lambda) \gg C_8(\Lambda)$, as ${\rm Im} \, C_7 \simeq
{\rm Im } \, C_8 \lesssim 2 \cdot 10^{-3}$.  We study the
corresponding CP asymmetries for $D \to PP \gamma$ in the Dalitz
region as this avoids large cancellations from $t$- or $u$-channel
resonances.  Note that the latter have not been included in
Ref.~\cite{Isidori:2012yx}.  We find values of $ A_{\rm CP}(s,t)$ up
to $\sim (3-4) \times 10^{-3}$ which is more than one order of
magnitude above the SM with maximal values of $\sim 2 \times 10^{-4}$,
shown in Fig.~\ref{fig:st-SM} for $D \to K^+ K^- \gamma$.  (As already
discussed, the corresponding SM asymmetry for $D \to \pi^+
\pi^-\gamma$ vanishes at this order.) The largest values for $A_{\rm
  CP}(s,t)$ arise around the resonances, notably the $K^* \to K
\gamma$ contributions to $D \to K K \gamma$.

The BSM CP asymmetries scale linearly with ${\rm Im} \,
C_7^{(\prime)}$. We checked explicitly that the CP asymmetries for
${\rm Im} \, C_7^{(\prime)} \simeq 2 \cdot 10^{-3}$ agree, up to an
overall suppression factor of 50, with those shown in
Fig.~\ref{fig:d2A_CP} which are based on ${\rm Im} \, C_7^{(\prime)}
\simeq 0.1$, and are therefore not shown.

Note that the $\Delta A_{\rm CP}$ constraint can be eased with a
strong phase suppression. In general, it can be escaped in the
presence of different sources of BSM CP violation in the hadronic
amplitudes. Yet, our analysis has shown that even with small CP
violation in the dipole couplings sizable NP enhancements can occur.

\section{Conclusions \label{sec:con}}

We worked out predictions for $D \to PP \gamma$ decay rates and
asymmetries in QCDF and in HH$\chi$PT. The $D \to \pi^+ \pi^- \gamma$
and $D \to K^+ K^- \gamma$ decays are sensitive to BSM physics, while
$D \to K \pi \gamma$ decays are SM-like and serve as ``standard
candles''. Therefore, a future measurement of the $D \to K \pi \gamma$
decay spectra can diagnose the performance of the QCD tools. The
forward-backward asymmetry (\ref{eq:AFB}) is particularly useful as it
vanishes for amplitudes without $t$- or $u$-channel dependence; this
happens, for instance, in leading-order QCDF.  On the other hand, $t$-
or $u$-channel resonances are included within HH$\chi$PT, and give
rise to finite interference patterns, shown in Fig.~\ref{fig:A_FB_SM}.
Within QCDF, the value of $\tilde C/\lambda_D$ can be extracted from
the branching ratio.

While branching ratios of $D \to \pi^+ \pi^- \gamma$ can be affected
by NP, these effects will be difficult to discern due to the large
uncertainties. On the other hand, the SM can be cleanly probed with CP
asymmetries in the $D \to \pi^+ \pi^- \gamma$ and $D \to K^+ K^-
\gamma$ decays, which can be sizable, see
Figs.~\ref{fig:CP_asymmetrie_QCDF}
and~\ref{fig:CP_asymmetrie_HHchiPT}. We stress that the sensitivity of
the CP asymmetries is maximized by performing a Dalitz analysis or
applying suitable cuts in $t$ (see Fig.~\ref{fig:d2A_CP}), as
otherwise large cancellations occur. Values of the CP asymmetries
depend strongly on the cut in $s$ employed to remove the
bremsstrahlung contribution. The latter is SM-like and dominates the
branching ratios for small photon energies. The forward-backward
asymmetries also offer SM tests, see
Fig.~\ref{fig_FB_asymmetrie_HHCHIPT}, but requires prior consolidation
of resonance effects.

Radiative charm decays are well-suited for investigation at the $e^+
e^-$ flavor facilities Belle II~\cite{Kou:2018nap},
BES~III~\cite{Ablikim:2019hff}, and future $e^+ e^-$-colliders running
at the $Z$-pole~\cite{Abada:2019lih}.  Branching ratios for $D^0 \to
\pi^+ \pi^- \gamma$ and $D^0 \to K^+ K^- \gamma$ decays are of the
order $10^{-5}$, see Table \ref{tbl:branching_ratios}. With
fragmentation fraction $f(c \to D^0) \simeq 0.59$ and $c\bar c$
production rates of $550 \cdot 10^9$ (Fcc-ee) and $65 \cdot 10^9$
(Belle II with $50 {\rm ab}^{-1}$) \cite{Abada:2019lih} this gives $6
\cdot 10^{11}$ and $8 \cdot 10^{10}$ neutral $D$-mesons and sizeable
(unreconstructed) event rates of $6 \cdot 10^6$ and $8 \cdot 10^5$,
respectively.  Rates for the ``standard candles'' $D^0 \to \pi^+
K^-\gamma$ are one order of magnitude larger. We look forward to
future investigations.

\section*{Acknowledgements}

We thank Svetlana Fajfer and Anita Prapotnik Brdnik for communication.
N.A. is supported in part by the DAAD.

\appendix

\section{Parameters}
\label{app:Parameter}
The couplings, masses, branching ratios, total decay widths and the mean life time  are taken from the PDG \cite{Tanabashi:2018oca}.
The mass of the $\eta_8$ results from the Gell-Mann-Okubo (GMO) mass formula \cite{Okubo:1962, Gell-Mann:1964}
\begin{align*}
  m_{\eta_8} = \sqrt{\frac{4m_K^2 - m_\pi^2}{3}}  = 0.56929 \GeV\, .
\end{align*}
The CKM matrix elements are taken from the UTfit collaboration \cite{UTfit}
\begin{align*}
	V_{ud} &= 0.97431 \pm 0.00012,  \quad \quad 
	V_{us} = 0.22514 \pm 0.00055, \\
	V_{cd} &= (-0.22500 \pm 0.00054) \exp \left[ \im (0.0351 \pm 0.0010)^{\circ} \right], \\
	V_{cs} &= (0.97344 \pm 0.00012) \exp \left[ \im (-0.001880 \pm 0.000055)^{\circ} \right].
\end{align*}
The decay constant of the D-meson is given by the FLAG working group \cite{Aoki:2016frl}
\begin{align*}
	&f_D = (0.21215 \pm 0.00145)\GeV,
	&f_{D_s} = (0.24883 \pm 0.00127)\GeV,\\
	&f_K = (0.1556 \pm 0.0004)\GeV,
	&f_\pi = (0.1302 \pm 0.0014)\GeV.
\end{align*}
The $q\qbar-s\sbar$ mixing scheme \cite{Feldmann:1998vh} and $\chi$PT \cite{Leutwyler:1997yr} provide decay constants for $\eta_8$ and $\eta_0$
\begin{align*}
  f_{\eta_8} = \sqrt{\frac{4}{3}f_K^2 - \frac{1}{3}f_\pi^2} = (0.1632 \pm 0.0006)\GeV\, , \\
  f_{\eta_0} = \sqrt{\frac{2}{3}f_K^2 + \frac{1}{3}f_\pi^2} = (0.1476 \pm 0.0005)\GeV\, .
\end{align*}
These values are in agreement with values extracted from $\eta^{(\prime)} \to \gamma \gamma$ decays \cite{Feldmann:1998vh}
\begin{align*}
  f_{\eta_8} =  (0.164 \pm 0.006)\GeV\,, 
  f_{\eta_0} =  (0.152 \pm 0.004)\GeV\,.
\end{align*}
The decay constants of the vector mesons are given by \cite{f_vector1, f_vector2} (and references therein)
\begin{align*}
	&f_{\rho} = (0.213 \pm 0.005)\GeV 
  &f_{\omega} = (0.197 \pm 0.008) \GeV, \\
 	&f_{\Phi} = (0.233 \pm 0.004)\GeV,
	&f_{K^*} = (0.204 \pm 0.007) \GeV.
\end{align*}

\section{Form factors}

\subsection{Vacuum $ \to PP$ transition form factors}

\label{appendix:em_weak_form_factors}

\begin{figure}
  \centering
  \includegraphics[width=0.9\linewidth]{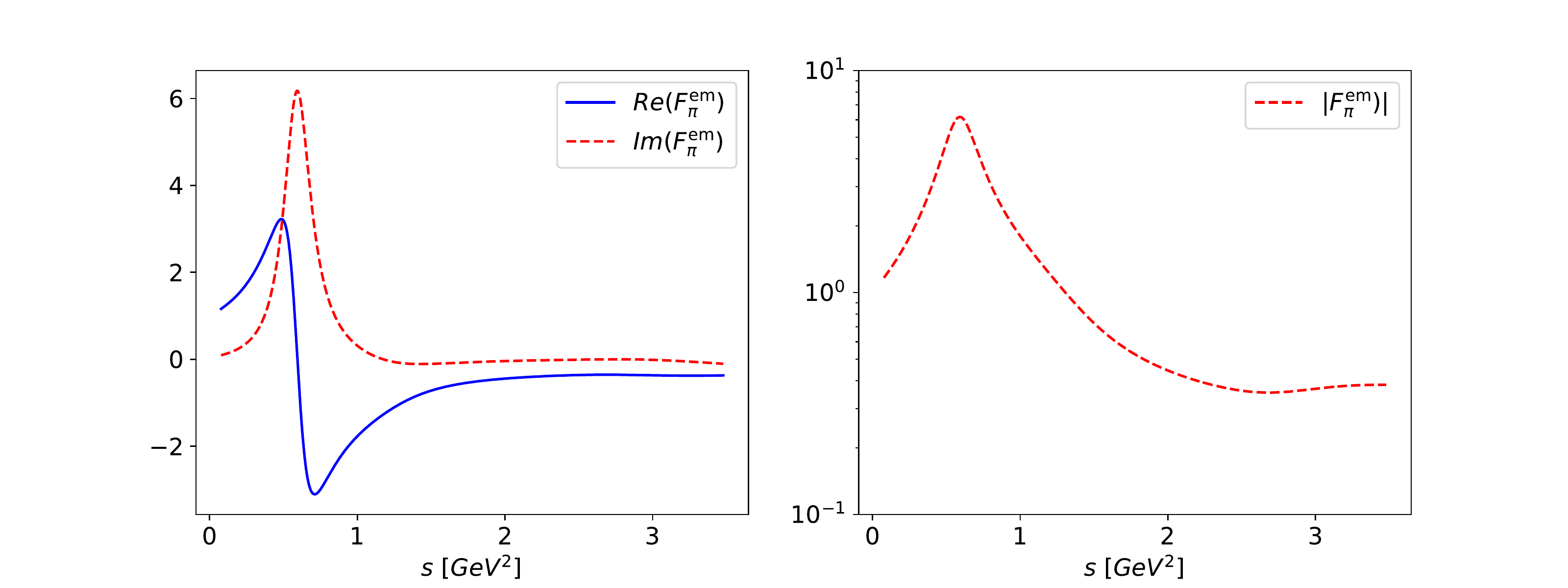}
  \caption{The real and imaginary part of the electromagnetic form factor $F_{\pi}^{\text{em}}$ (\ref{eq:Fpi}) (left) as well as the absolute value (right)  as a function of the invariant mass squared $s$.}
  \label{fig:em pion form factor}
\end{figure}

The electromagnetic pion form factor $F_{\pi}^{\text{em}}$ is  defined as
\begin{align}
	&\braket{\pi^+(p_1) \pi^-(p_2)|j^{\text{em}}_{\mu}|0} = (p_1 - p_2)_{\mu} F_{\pi}^{\text{em}}(s) \, , 
\end{align}
with 
the electromagnetic current
\begin{align}
  \begin{split}
    j_{\mu}^{\text{em}} &= \frac 23 \overline{u} \gamma_{\mu} u - \frac 13 \overline{d} \gamma_{\mu} d - \frac 13 \overline{s} \gamma_{\mu} s\\
	             &= \frac12 (\overline{u} \gamma_{\mu} u - \overline{d} \gamma_{\mu} d) + \frac16 (\overline{u} \gamma_{\mu} u + \overline{d} \gamma_{\mu} d) - \frac 13 \overline{s} \gamma_{\mu} s\\
               &= \frac{1}{\sqrt{2}}j_{\mu}^{(I=1)} + \frac{1}{3\sqrt{2}}j_{\mu}^{(I=0)} - \frac{1}{3} j_{\mu}^s \\
               &= J_{\mu}^{(I=1)} + J_{\mu}^{(I=0)} + J_{\mu}^s \, . 
  \end{split}
\end{align}
 In the isospin symmetry limit, only the $I=1$ current contributes to $F_{\pi}^{\text{em}}$,
which reads \cite{Bruch:2004py}
\begin{align} \label{eq:Fpi}
	F_{\pi}^{\text{em}}(s) = \left[ \sum_{n=0}^{3} c_n BW_n^{KS}(s)\right]_{fit} +  \left[ \sum_{n=4}^{\infty} c_n BW_n^{KS}(s)\right]_{dual-QCD_{N_C=\infty}},
\end{align}
where the coefficients $c_n$ are given by
\begin{align}
  \begin{split}
    c_0 &= 1.171\pm 0.007, \quad c_1 = -0.119\pm 0.011, \\
	c_2 &= 0.0115 \pm 0.0064, \quad c_3 = -0.0438\pm 0.02,\\
	c_n &= \frac{2(-1)^n \Gamma(1.8)m_{\rho}^2}{\sqrt{\pi} m_n^2 \Gamma(n+1)\Gamma(1.3-n)} \qquad n\geq 4, \\
	m_n^2 &= m_{\rho}^2(1+2n)
  \end{split}
\end{align}
and the functions $BW_n^{KS}(s)$ read
\begin{align}
  \begin{split}
    BW_n^{KS}(s) &= \frac{m_n^2}{m_n^2 - s - \im \sqrt{s} \Gamma_n(s)},\\
	\Gamma_n(s)&= \frac{0.2m_n^3}{s} \left( \frac{p(s)}{p(m_n^2)} \right)^3,\\
	p(s) &= 0.5 \sqrt{s-4m_{\pi}^2}.
  \end{split}
\end{align}
The masses and widths of the $\rho$ meson and its first resonance are fitted as well
\begin{align}
  \begin{split}
    &m_{\rho} = (0.7739\pm 0.0006)\GeV, \quad m_{\rho'} = (1.357 \pm 0.018)\GeV,\\
    &\Gamma_{\rho}= (0.1149 \pm 0.0010)\GeV, \quad \Gamma_{\rho'}=(0.437\pm 0.060)\GeV.
  \end{split}
\end{align}
$F_{\pi}^{\text{em}}$  is  shown in Figure \ref{fig:em pion form factor}.

\begin{figure}
  \centering
  \includegraphics[width=0.9\linewidth]{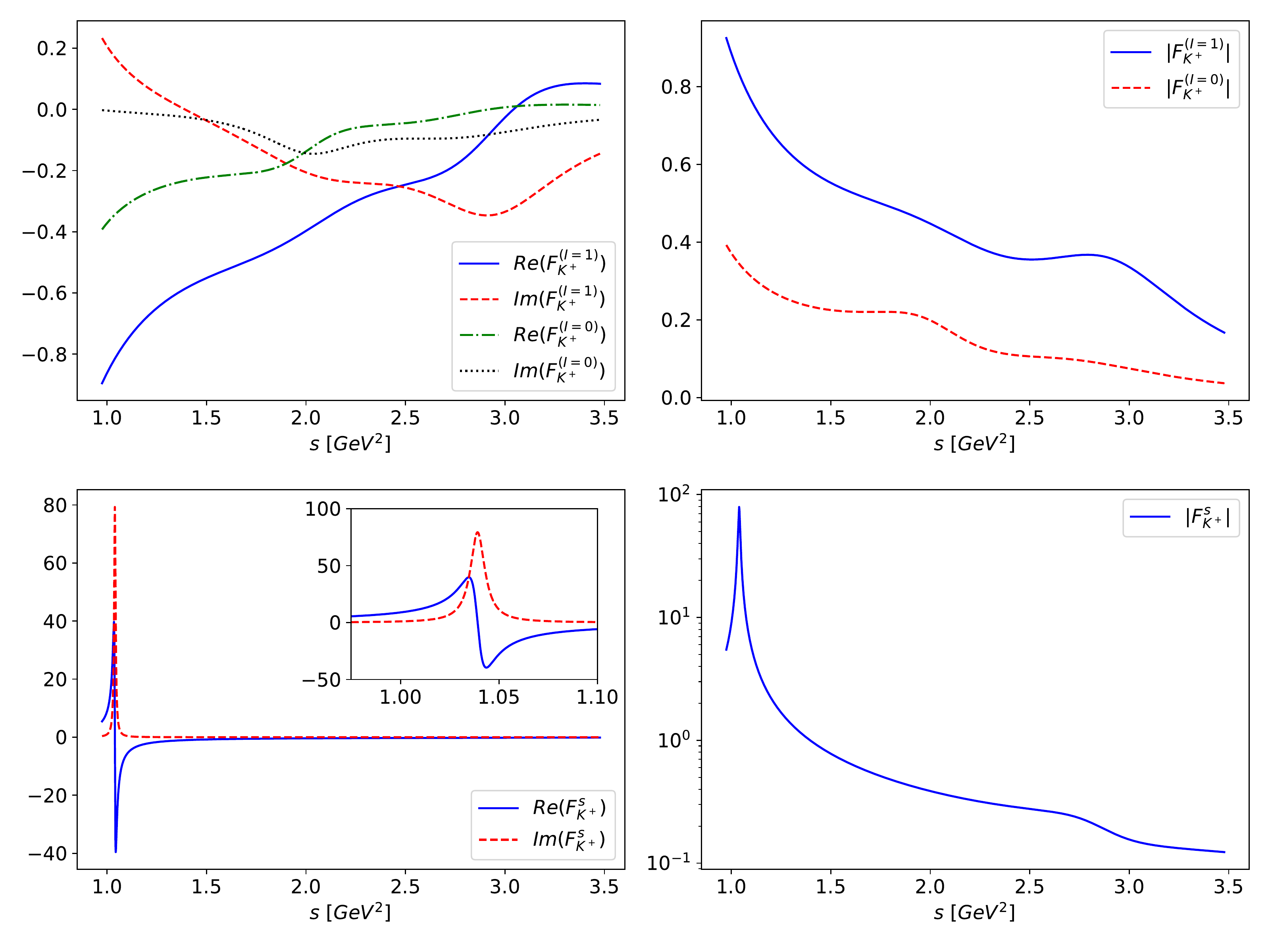}
  \caption{The real and imaginary parts of the electromagnetic kaon form factors (\ref{eq:FK}) (left) as well as their absolute values (right) as a function of $s$. The upper (lower) plots show $F_{K^+}^{(I=1,0)}$ ($F_{K^+}^s$).}
  \label{fig:em kaon form factor}
\end{figure}

The electromagnetic kaon form factor $F_{K^+}^{\text{em}}$, defined as
\begin{align}
	\braket{K^+(p_1) K^-(p_2)|j^{\text{em}}_{\mu}|0} = (p_1 - p_2)_{\mu} F_{K^+}^{\text{em}}(s) \, , 
\end{align}
is taken from \cite{Bruch:2004py} and shown in Figure \ref{fig:em kaon form factor}. It can be decomposed into an isospin-one component $F_{K^+}^{(I=1)}$ and two isospin-zero components
$F_{K^+}^{(I=0)}$,  $F_{K^+}^s$, with $\omega$ and $\phi$ contributions, respectively,
\begin{align} \label{eq:FK}
  \begin{split}
    F_{K^+}^{\text{em}}(s)    &= F_{K^+}^{(I=1)}(s) + F_{K^+}^{(I=0)}(s) + F_{K^+}^s(s),\\
	F_{K^+}^{(I=1)}(s) &= \frac 1 2 (c_{\rho}^K BW_{\rho}(s) + c_{\rho'}^K BW_{\rho'}(s) + c_{\rho''}^K BW_{\rho''}(s)), \\
	F_{K^+}^{(I=0)}(s) &= \frac 1 6 (c_{\omega}^K BW_{\omega}(s) + c_{\omega'}^K BW_{\omega'}(s) + c_{\omega''}^K BW_{\omega''}(s)) , \\
	F_{K^+}^s(s)       &= \frac 1 3 (c_{\phi} BW_{\phi}(s) + c_{\phi'} BW_{\phi'}(s)).
  \end{split}
\end{align}
The requisite parameters are given by
\begin{align}
  \begin{split}
    &m_{\phi} = 1.019372 \GeV, \quad m_{\phi} = 1.68\GeV, \quad m_{\rho'} = 1.465\GeV,\\
	  &m_{\rho''} = 1.720\GeV, \quad m_{\omega'} = 1.425\GeV, \quad m_{\omega''} = 1.67\GeV,\\
	  &\Gamma_{\phi} = 0.00436\GeV, \quad \Gamma_{\phi'} = 0.150\GeV, \quad \Gamma_{\rho} = 0.150\GeV, \quad \Gamma_{\rho'} = 0.400\GeV,\\
	  &\Gamma_{\rho''} = 0.250\GeV, \quad \Gamma_{\omega} = 0.0084\GeV, \quad \Gamma_{\omega'} = 0.215\GeV, \quad \Gamma_{\omega''} = 0.315\GeV,\\
	  &c_{\phi} = (1.018\pm 0.006), \quad c_{\phi'} = (-0.018\pm 0.006), \\
	  & c^K_{\rho} = (1.195\pm0.009), \quad c^K_{\rho'} = (-0.112\pm0.010), \quad c^K_{\rho''} = (-0.083\pm0.019), \\
	  &c^K_{\omega} = (1.195\pm0.009), \quad c^K_{\omega'} = (-0.112\pm0.010), \quad c^K_{\omega''} = (-0.083\pm0.019).
  \end{split}
\end{align}

\begin{figure}
  \centering
  \includegraphics[width=0.9\linewidth]{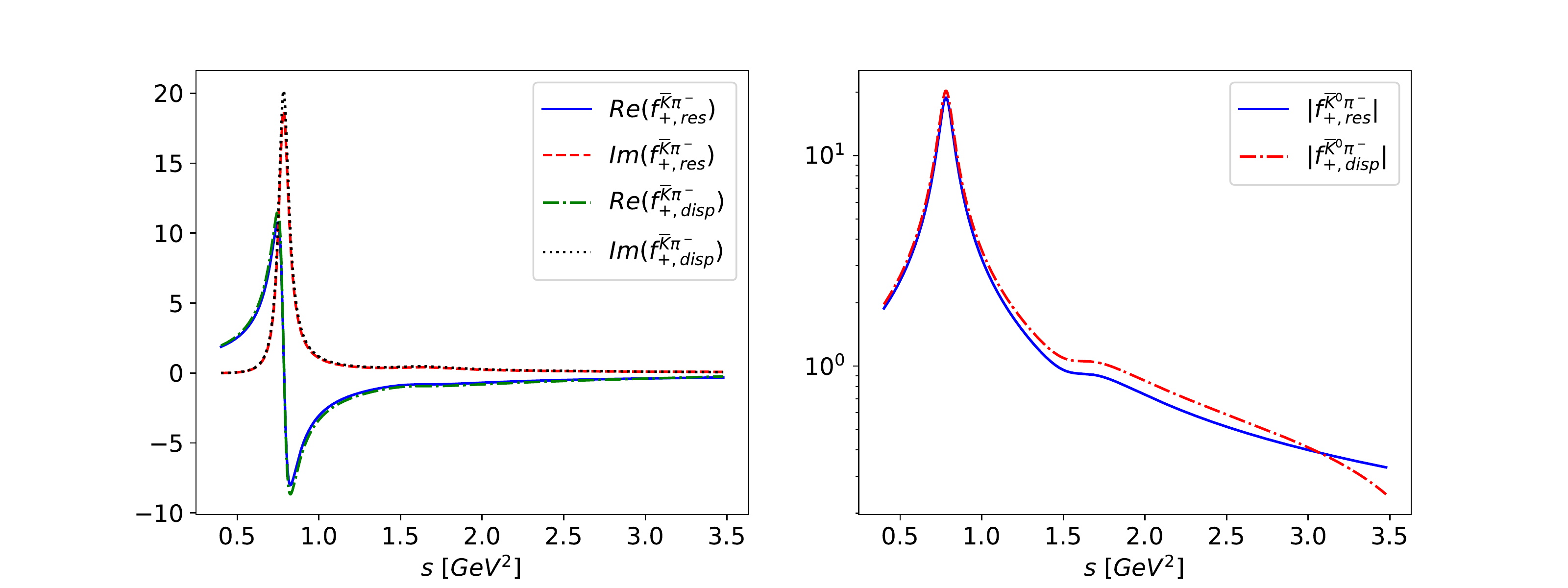}
  \caption{The real and imaginary part (left) of the $f_+^{\overline{K}\pi^-}$ form factor (\ref{eq:FKpi}) as well as the absolute value (right) versus $s$ in  the two resonance models as well as in the dispersive description. The form factor is extracted from $\tau^- \rightarrow \nu_\tau K_s \pi^-$ decays \cite{Boito:2008fq}.
  For $\bar K^0 \pi^0$ and $K^+ \pi^-$, we use isospin relations (\ref{eq:iso}).}
  \label{fig:Kpi form factor}
\end{figure}

The $\overline{K} \pi^-$ form factors are defined as
\begin{align}  \label{eq:FKpi}
    \braket{\pi^-(p_1) \overline{K}(p_2)| \sbar \gamma_\mu u| 0} &= f_+^{\overline{K}\pi^-}(s) \left(p_2 - p_1\right)_\mu + f_-^{\overline{K}\pi^-}(s) \left(p_2 + p_1\right)_\mu\\
    &= -\frac{\Delta_{K \pi}}{s} f_0^{\overline{K}\pi^-}(s) \left(p_2 + p_1\right)_\mu + \left[ \left(p_2 - p_1\right)_\mu + \frac{\Delta_{K \pi}}{s}  \left(p_2 + p_1\right)_\mu\right] f_+^{\overline{K}\pi^-}(s)\, , \nonumber
\end{align} 
with 
$\Delta_{K\pi} = m_K^2-m_\pi^2$. The vector form factor $f_+^{\overline{K}\pi^-}$,  shown in Figure \ref{fig:Kpi form factor},  can be parametrized with a dispersion relation with three subtractions  at $s=0$ \cite{Boito:2008fq} 
\begin{align}
  f_+^{\overline{K} \pi^-}(s) = f_+^{\overline{K} \pi^-}(0) \cdot \exp \left[\lambda_+' \frac{s}{m_\pi^2} + \frac{1}{2} \left(\lambda_+'' - \lambda_+'^2\right)\frac{s}{m_\pi^4} + \frac{s^3}{\pi} \int_{s_{K\pi}}^{s_{cut}} \dd{}s' \frac{\delta_1^{K\pi}(s')}{(s')^3(s'-s-\im \epsilon)}\right]\, ,
\end{align}
with $s_{K\pi}=(m_K + m_\pi)^2$. The phase $\delta_1^{K\pi}(s)$ is extracted from a two resonance model \cite{Boito:2008fq}
\begin{align}
  \tilde{f}_+^{\overline{K}\pi^-}(s) = \frac{f_+^{\overline{K}\pi^-}(s)}{f_+^{\overline{K}\pi^-}(0)} = \frac{m^2_{K^\star} - \kappa_{K^\star} \tilde{H}_{K\pi}(0) + \beta s}{D(K^\star)} - \frac{\beta s}{D(K^{\star \prime})}\, ,
\end{align}
where
\begin{align}
  \begin{split}
   &  D(n) = m_n^2 - s - \kappa_n \text{Re}\left(\tilde{H}_{K\pi}(s)\right) - \im m_n \gamma_n(s)\, , \\
    &\gamma_n(s) = \gamma_n \frac{s}{m_n^2} \frac{\sigma_{K\pi}^3(s)}{\sigma_{K\pi}^3(m_n^2)}, \qquad \gamma_n = \gamma_n(m_n^2)\, ,\\
    &\sigma_{K\pi}(s) = \frac{2q_{K\pi}(s)}{\sqrt{s}} = \frac{1}{s} \sqrt{\left(s - (m_K + m_\pi)^2\right)\left(s - (m_K - m_\pi)^2\right)}\, ,\\
    &\kappa_n = \frac{192\pi f_K f_\pi}{\sigma^3_{K\pi}(m_n^2)}\frac{\gamma_n}{m_n}\, .
  \end{split}
\end{align}
The function $\tilde{H}_{K\pi}$ is a $\chi$PT loop integral function~\cite{Jamin:2006tk}
\begin{align}
  \tilde{H}_{K\pi}(s) = H_{K\pi}(s) - \frac{2}{3 f_\pi^2}L_{K\pi}^r s = \frac{1}{f_\pi^2}\left[s M_{K\pi}^r(s) - L_{K\pi}(s)\right]\, ;
\end{align}
explicit expressions for $M^r(s)$ and $L(s)$ can be found in chapter 8 of Ref.~\cite{Gasser1985}:
\begin{align}
  \begin{split}
    &M_{K\pi}^r(s) = \frac{1}{12s}(s-2\Sigma)\overline{J}_{K\pi}(s) + \frac{\Delta^2}{3s^2}\overline{\overline{J}}_{K\pi}(s) - \frac{1}{6}k_{K\pi}(\mu) + \frac{1}{288\pi^2}\, ,\\
    &L_{K\pi}(s)= \frac{\Delta^2}{4s}\overline{J}_{K\pi}(s)\, ,\\
    &k_{K\pi}(\mu) = \frac{1}{32\pi^2} \frac{1}{\Delta} \left(m_K^2 \ln\left(\frac{m_K^2}{\mu^2}\right) - m_\pi^2 \ln\left(\frac{m_\pi^2}{\mu^2}\right)\right)\, , \\
    &\overline{\overline{J}}_{K\pi}(s) = \overline{J}_{K\pi}(s) - s \overline{J}_{K\pi}^\prime(0)\, ,\\
    &\overline{J}_{K\pi}(s) = J_{K\pi}(s) - J_{K\pi}(0) \\
    &\qquad \quad = \frac{1}{32\pi^2} \left(2 + \left[\frac{\Delta}{s} - \frac{\Sigma}{\Delta}\right]\ln\left(\frac{m_\pi^2}{m_K^2}\right) - \frac{v}{s}\ln\left(\frac{(s+v)^2 - \Delta^2}{(s-v)^2 - \Delta^2}\right)\right)\, , \\
    &\overline{J}_{K\pi}^\prime(0) = \frac{1}{32\pi^2} \left(\frac{\Sigma}{\Delta^2} + 2\frac{m_K^2 m_\pi^2}{\Delta^3}\ln\left(\frac{m_\pi^2}{m_K^2}\right)\right)\, , \\
    &v(s) = s\sigma_{K\pi}(s)\, , \\
    &\Sigma = m_K^2 + m_\pi^2\, ,\\
    &\Delta = m_K^2 - m_\pi^2\, .
  \end{split} 
\end{align}
The renormalization scale $\mu$ is set to the physical resonance mass $\mu = m_{K^\star}$ \cite{Boito:2008fq}. The resonance masses and width parameters are unphysical fitting parameters. They are obtained as \cite{Boito:2008fq}
\begin{align}
  \begin{split}
    &m_{K^\star}^{\text{fit}} = (0.94341 \pm 0.00058) \GeV\, , \qquad \gamma_{K^\star}^{\text{fit}} = (0.06672 \pm 0.00086) \GeV\, , \\
    &m_{K^{\star\prime}}^{\text{fit}} = (1.374 \pm 0.030) \GeV\, , \qquad \gamma_{K^{\star\prime}}^{\text{fit}} = (0.24 \pm 0.10) \GeV\, , \\
    &s_{cut} = 4\GeV^2\, , \qquad \mu = m^{\text{phy}}_{K\star} =  0.892 \GeV\, , \qquad \beta =(-3.9\pm 1.5)\cdot10^{-2}\, ,\\
    &\lambda_+^\prime=(24.66\pm 0.69)\cdot10^{-3}\, , \qquad \lambda_+^{\prime \prime}=(11.99\pm 0.19)\cdot10^{-4}\, ,\\
    &|V_{us}|f_+^{\overline{K}\pi^-}(0)= 0.21664 \pm 0.00048 \, . 
  \end{split}
\end{align} 

\subsection{HH$\chi$PT form factors}
\label{app:HQCHPT form factors}
\subsubsection{Vector form factors}
\begin{figure}
  \centering
  \includegraphics[width=0.7\linewidth]{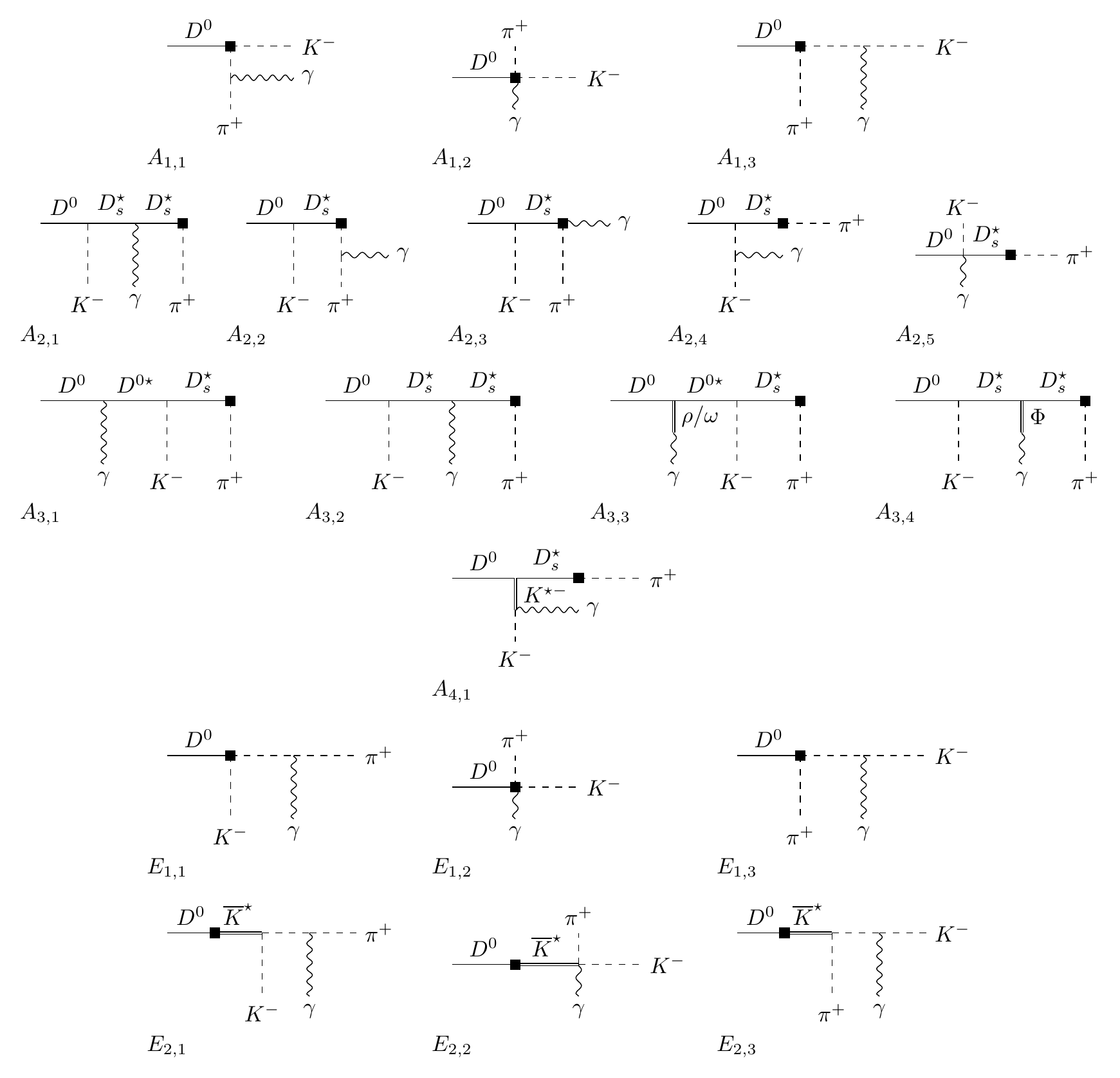}
  \caption{Feynman diagrams for the $D\to \pi^+ K^- \gamma$ decay, which contribute to the parity-even form factors $A$ and $E$. The diagrams for the decays $D\to \pi^+ \pi^- \gamma$ and $D\to K^+ K^- \gamma$ are obtained by adjusting the flavors. We have added the diagrams $E_{1,2}$ and $E_{2,2}$ (see \cite{Fajfer:2002bq}) to make the amplitude E gauge invariant for any choice of a. Additionally, for each of the diagrams $A_{1,1}$, $A_{1,2}$, $A_{1,3}$, $A_{2,2}$, $A_{2,3}$, $A_{2,4}$, $E_{1,1}$, $E_{1,2}$, $E_{1,3}$, $E_{2,1}$ and $E_{2,3}$ there is another one where the photon is coupled via a vector meson.}
  \label{fig:Diagramme_AE}
\end{figure}
\begin{figure}
  \centering
  \includegraphics[width=0.7\linewidth]{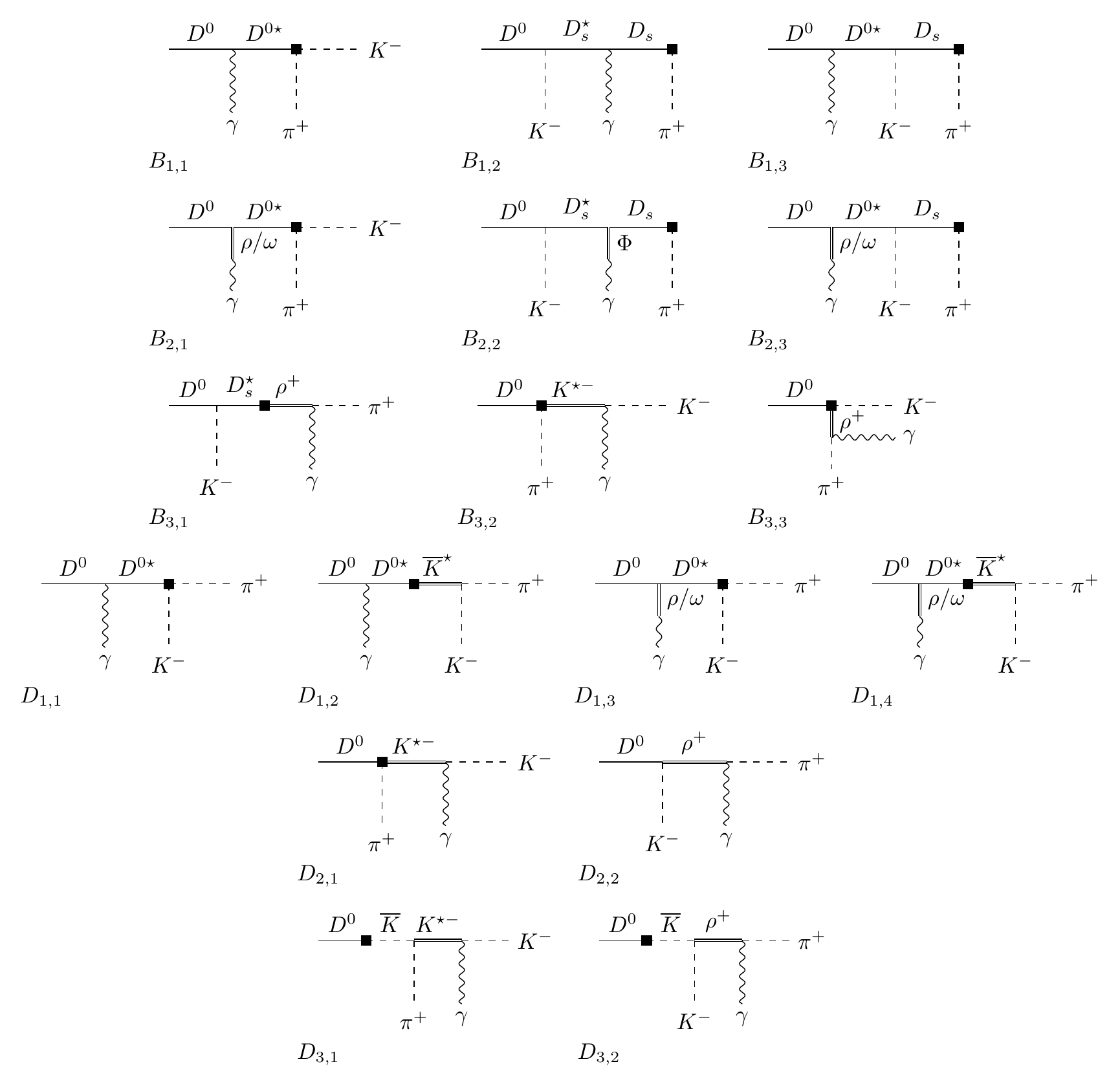}
  \caption{Feynman diagrams for the $D\to \pi^+ K^- \gamma$ decay, which contribute to the parity-odd form factors $B$ and $D$. The diagrams for the decays $D\to \pi^+ \pi^- \gamma$ and $D\to K^+ K^- \gamma$ are obtained by adjusting the flavors.}
  \label{fig:Diagramme_BD}
\end{figure}
\noindent\underline{$D \to \pi^0 \overline{K} \gamma$}
\begin{align}
  \begin{split}
    E_1^{(d,s)} &= \im  g \frac{f_D f_K}{f_\pi} \frac{v\cdot k}{v\cdot k + v\cdot p_1 + \Delta}\left(\frac{1}{v\cdot k + \Delta} - \frac{1}{v\cdot p_1 + \Delta}\right) \left(\sqrt{2}\lambda^\prime + \frac{1}{2} \lambda g_v \left(\frac{g_\omega}{3m_\omega^2} + \frac{g_\rho}{m_\rho^2}\right)\right)
  \end{split}
\end{align}
\begin{align}
  \begin{split}
    D_1^{(d,s)} &= -\sqrt{2} \frac{f_D f_K}{f_\pi}\lambda^\prime \left(\frac{1}{v\cdot k + \Delta} + g \frac{v\cdot p_2}{v\cdot p_1 + v\cdot k}\left[\frac{1}{v\cdot k + \Delta}+\frac{1}{v\cdot p_1 + \Delta}\right]\right)\\
    D_2^{(d,s)} &= -\frac{1}{2}\frac{f_D f_K}{f_\pi} \lambda g_v \left(\frac{g_\omega}{3m_\omega^2} + \frac{g_\rho}{m_\rho^2}\right)\left(\frac{1}{v\cdot k + \Delta} +g \frac{v\cdot p_2}{v\cdot p_1 + v\cdot k}\left[\frac{1}{v\cdot k + \Delta}+\frac{1}{v\cdot p_1 + \Delta}\right]\right)\\
    D_3^{(d,s)} &= -\sqrt{\frac{2}{m_D}} f_K (\alpha_1 m_D - \alpha_2 v\cdot p_2)\left(\frac{m_\rho^2g_{\rho \pi \gamma}}{g_\rho} BW_\rho(p_1+k) + \frac{m_\omega^2g_{\omega \pi \gamma}}{g_\omega} BW_\omega(p_1+k)\right)\\
                &+ \frac{1}{\sqrt{2}}g_{\overline{K}^\star} g_{\overline{K}^\star \overline{K} \gamma} \frac{f_D}{f_\pi} \left(1+ g\frac{m_D-v\cdot p_1}{v\cdot p_1 + \Delta}\right) BW_{\overline{K}^\star}(p_2+k)\\
    D_4^{(d,s)} &=-f_D \frac{1}{v\cdot k + \Delta} \left(1 - m_{K^\star}^2 BW_{\overline{K}^\star}(p_1 + p_2)\right)\left(\sqrt{2}\lambda^\prime + \frac{1}{2} \lambda g_v \left(\frac{g_\omega}{3m_\omega^2} + \frac{g_\rho}{m_\rho^2}\right)\right)\\
    D_5^{(d,s)} &=\frac{1}{\sqrt{2}}\frac{f_D}{f_K} (g_\rho g_{\rho \pi \gamma} BW_\rho(p_1+k) - g_\omega g_{\omega \pi \gamma} BW_\omega(p_1+k))\\
    D_6^{(d,s)} &= \frac{1}{\sqrt{2}}f_D f_K \frac{m_D^2}{m_D^2 - m_{K^0}^2} \left(\frac{m_\rho^2}{g_\rho}g_{\rho \pi \gamma} BW_\rho(p_1+k) - \frac{m_\omega^2}{g_\omega}g_{\omega \pi \gamma} BW_\omega(p_1+k)\right)
  \end{split}
\end{align}

\noindent\underline{$D \to \pi^+ K^- \gamma$}

\begin{align}
  \begin{split}
    A_1^{(d,s)} &= \im \frac{f_D f_\pi}{f_K}\frac{p_1\cdot k - m_D(v\cdot k + v\cdot p_1)}{(p_1\cdot k)(p_2\cdot k)} \\
    A_2^{(d,s)} &= -\im \sqrt{m_Dm_{D_s}} \frac{f_{D_s}f_\pi}{f_K} \frac{g}{(p_1\cdot k)(p_2\cdot k)} \left(\frac{p_2\cdot k(m_D - v\cdot p_2)}{m_D(v\cdot p_2 +\Delta)} + \frac{p_2\cdot k \left[p_1\cdot p_2 - (v\cdot p_1)(v\cdot p_2)\right]}{m_D(v\cdot p_2+\Delta)(v\cdot p_2 + v\cdot k + \Delta)} \right. \\
                &+ \left. \frac{(m_D-v\cdot p_1)p_1\cdot k + m_Dp_1\cdot p_2 + m_D(v\cdot p_1)(v\cdot p_2)}{m_D(v\cdot p_2 + v\cdot k + \Delta)}\right) \\
    A_3^{(d,s)} &= \im \sqrt{\frac{m_{D_s}}{m_D}} \frac{f_{D_s}f_\pi}{f_K} g \frac{v\cdot k}{v\cdot k + v\cdot p_2 + \Delta}\left(\frac{2\lambda^\prime - \frac{\sqrt{2}}{3}\lambda g_v \frac{g_\Phi}{m_\Phi^2}}{v\cdot p_2 + \Delta} - \frac{2\lambda^\prime + \frac{1}{\sqrt{2}}\lambda g_v \left(\frac{g_\omega}{3m_\omega^2} + \frac{g_\rho}{m_\rho^2}\right)}{v\cdot k + \Delta}\right)
  \end{split}
\end{align}
\begin{align} \label{eq:BpiK}
  \begin{split}
    B_1^{(d,s)} &= 2 \frac{f_D f_\pi}{f_K}\lambda^\prime \left(\frac{1}{v\cdot k + \Delta} + g \frac{f_{D_s}}{f_D }\sqrt{\frac{m_{D_s}}{m_D}}\frac{v\cdot p_1}{(v\cdot p_2 + v\cdot k)}\left[\frac{1}{v\cdot k + \Delta}+\frac{1}{v\cdot p_2 + \Delta}\right]\right)\\
    B_2^{(d,s)} &= \frac{1}{\sqrt{2}}\frac{f_\pi}{f_K} \lambda g_v \left(\frac{g_\omega}{3m_\omega^2} + \frac{g_\rho}{m_\rho^2}\right)\frac{1}{v\cdot k + \Delta}\left(f_D + g \sqrt{\frac{m_{D_s}}{m_D}}f_{D_s} \frac{v\cdot p_1}{(v\cdot p_2 + v\cdot k)}\right)\\
                &- g \lambda g_v \sqrt{\frac{m_{D_s}}{m_D}}\frac{f_{D_s} f_\pi}{f_K} \frac{\sqrt{2}g_\Phi}{3m_\Phi^2}\frac{v\cdot p_1}{(v\cdot p_2 + v\cdot k)(v\cdot p_2 + \Delta)}\\
    B_3^{(d,s)} &= \frac{2}{\sqrt{m_D}} f_\pi\frac{m^2_{K^{\star}}g_{K^{\star\pm} K^\pm \gamma}}{g_{K^{\star}}}  (\alpha_1 m_D - \alpha_2 v\cdot p_1)BW_{K^{\star-}}(p_2+k)\\
                &- g_{\rho} g_{\rho^\pm \pi^\pm \gamma} \frac{f_D}{f_K} \left(1+ g\sqrt{\frac{m_{D_s}}{m_D}}\frac{f_{D_s}}{f_D}\frac{m_D-v\cdot p_2}{v\cdot p_2 + \Delta}\right) BW_{\rho^+}(p_1+k)
  \end{split}
\end{align}
\begin{align}
  \begin{split}
    D_1^{(d,s)} &= \frac{f_D}{v\cdot k + \Delta} \left(1 - m_{K^\star}^2 BW_{\overline{K}^\star}(p_1 + p_2)\right)\left(2\lambda^\prime + \frac{1}{\sqrt{2}} \lambda g_v \left(\frac{g_\omega}{3m_\omega^2} + \frac{g_\rho}{m_\rho^2}\right)\right)\\
    D_2^{(d,s)} &=-\frac{f_D}{f_\pi} g_{K^\star} g_{K^{\star\pm} K^\pm \gamma} BW_{K^{\star-}}(p_2+k) - \frac{f_D}{f_K} g_\rho g_{\rho^\pm \pi^\pm \gamma} BW_{\rho^+}(p_1+k)\\
    D_3^{(d,s)} &= {f_D f_K} \frac{m_D^2}{m_D^2 - m_{K^0}^2} \left(\frac{m_\rho^2}{g_\rho}g_{\rho^\pm \pi^\pm \gamma} BW_{\rho^+}(p_1+k) +\frac{m_{K^\star}^2}{g_{K^\star}}g_{K^{\star\pm} K^\pm \gamma} BW_{K^{\star-}}(p_2+k)\right)
  \end{split}
\end{align}

\noindent\underline{$D \to \pi^+ \pi^- \gamma$}

\begin{align}
  \begin{split}
    A_1^{(d,d)} &= \im f_D\frac{p_1\cdot k - m_D(v\cdot k + v\cdot p_1)}{(p_1\cdot k)(p_2\cdot k)} \\
    A_2^{(d,d)} &= -\im m_D f_{D} \frac{g}{(p_1\cdot k)(p_2\cdot k)} \left(\frac{p_2\cdot k(m_D - v\cdot p_2)}{m_D(v\cdot p_2 +\Delta)} + \frac{p_2\cdot k \left[p_1\cdot p_2 - (v\cdot p_1)(v\cdot p_2)\right]}{m_D(v\cdot p_2+\Delta)(v\cdot p_2 + v\cdot k + \Delta)} \right. \\
    &+ \left. \frac{(m_D-v\cdot p_1)p_1\cdot k + m_Dp_1\cdot p_2 + m_D(v\cdot p_1)(v\cdot p_2)}{m_D(v\cdot p_2 + v\cdot k + \Delta)}\right) \\
    A_3^{(d,d)} &= \im f_D g \frac{v\cdot k}{v\cdot k + v\cdot p_2 + \Delta}\left(\frac{2\lambda^\prime + \frac{1}{\sqrt{2}}\lambda g_v \left(\frac{g_\omega}{3m_\omega^2} - \frac{g_\rho}{m_\rho^2}\right)}{v\cdot p_2 + \Delta} - \frac{2\lambda^\prime + \frac{1}{\sqrt{2}}\lambda g_v \left(\frac{g_\omega}{3m_\omega^2} + \frac{g_\rho}{m_\rho^2}\right)}{v\cdot k + \Delta}\right) 
  \end{split}
\end{align}


\begin{align} \label{eq:Bpipi}
  \begin{split}
    B_1^{(d,d)} &= 2 f_D\lambda^\prime \left(\frac{1}{v\cdot k + \Delta} + g \frac{v\cdot p_1}{(v\cdot p_2 + v\cdot k)}\left[\frac{1}{v\cdot k + \Delta}+\frac{1}{v\cdot p_2 + \Delta}\right]\right)\\
    B_2^{(d,d)} &= \frac{1}{\sqrt{2}} \lambda g_v f_D \left(\frac{g_\omega}{3m_\omega^2} + \frac{g_\rho}{m_\rho^2}\right)\frac{1}{v\cdot k + \Delta}\left(1 + g \frac{v\cdot p_1}{(v\cdot p_2 + v\cdot k)}\right)\\
                &+ \frac{1}{\sqrt{2}}g \lambda g_v f_D \left(\frac{g_\omega}{3m_\omega^2} - \frac{g_\rho}{m_\rho^2}\right)\frac{v\cdot p_1}{(v\cdot p_2 + v\cdot k)(v\cdot p_2 + \Delta)}\\
    B_3^{(d,d)} &= \frac{2}{\sqrt{m_D}} f_\pi \frac{m_\rho^2}{g_\rho}g_{\rho^{\pm} \pi^\pm \gamma} (\alpha_1 m_D - \alpha_2 v\cdot p_1)BW_{\rho^-}(p_2+k)\\
                &-  g_{\rho} g_{\rho^\pm \pi^\pm \gamma} \frac{f_D}{f_\pi} \left(1+ g\frac{m_D-v\cdot p_2}{v\cdot p_2 + \Delta}\right) BW_{\rho^+}(p_1+k)
  \end{split}
\end{align}

\begin{align}
  \begin{split}
    D_1^{(d,d)} &=f_D \frac{1}{v\cdot k + \Delta} \left(1 - m_\rho^2 BW_{\rho}(p_1 + p_2)\right)\left(2\lambda^\prime + \frac{1}{\sqrt{2}} \lambda g_v \left(\frac{g_\omega}{3m_\omega^2} + \frac{g_\rho}{m_\rho^2}\right)\right)\\
    D_2^{(d,d)} &= -\frac{f_D}{f_\pi} g_{\rho} g_{\rho^{\star\pm} \pi^\pm \gamma} (BW_{\rho^-}(p_2+k) +  BW_{\rho^+}(p_1+k))\\
    D_3^{(d,d)} &= {f_D f_\pi } \frac{m_D^2}{m_D^2 - m_{\pi^0}^2} \frac{m_\rho^2}{g_\rho}g_{\rho^\pm \pi^\pm \gamma} \left(BW_{\rho^+}(p_1+k) + BW_{\rho^-}(p_2+k)\right)
  \end{split}
\end{align}

\noindent\underline{$D \to K^+ K^- \gamma$}

\begin{align}
  \begin{split}
    A_1^{(s,s)} &= \im f_D\frac{p_1\cdot k - m_D(v\cdot k + v\cdot p_1)}{(p_1\cdot k)(p_2\cdot k)} \\
    A_2^{(s,s)} &= -\im \sqrt{m_Dm_{D_s}} f_{D_s} \frac{g}{(p_1\cdot k)(p_2\cdot k)} \left(\frac{p_2\cdot k(m_D - v\cdot p_2)}{m_D(v\cdot p_2 +\Delta)} + \frac{p_2\cdot k \left[p_1\cdot p_2 - (v\cdot p_1)(v\cdot p_2)\right]}{m_D(v\cdot p_2+\Delta)(v\cdot p_2 + v\cdot k + \Delta)} \right. \\
    &+ \left. \frac{(m_D-v\cdot p_1)p_1\cdot k + m_Dp_1\cdot p_2 + m_D(v\cdot p_1)(v\cdot p_2)}{m_D(v\cdot p_2 + v\cdot k + \Delta)}\right) \\
    A_3^{(s,s)} &= \im \sqrt{\frac{m_{D_s}}{m_D}} f_{D_s} g \frac{v\cdot k}{v\cdot k + v\cdot p_2 + \Delta}\left(\frac{2\lambda^\prime - \frac{\sqrt{2}}{3}\lambda g_v \frac{g_\Phi}{m_\Phi^2}}{v\cdot p_2 + \Delta} - \frac{2\lambda^\prime + \frac{1}{\sqrt{2}}\lambda g_v \left(\frac{g_\omega}{3m_\omega^2} + \frac{g_\rho}{m_\rho^2}\right)}{v\cdot k + \Delta}\right) 
  \end{split}
\end{align}
\begin{align} \label{eq:BKK}
  \begin{split}
    B_1^{(s,s)} &= 2 f_D\lambda^\prime \left(\frac{1}{v\cdot k + \Delta} + g \frac{f_{D_s}}{f_D }\sqrt{\frac{m_{D_s}}{m_D}}\frac{v\cdot p_1}{(v\cdot p_2 + v\cdot k)}\left[\frac{1}{v\cdot k + \Delta}+\frac{1}{v\cdot p_2 + \Delta}\right]\right)\\
    B_2^{(s,s)} &= \frac{1}{\sqrt{2}}\lambda g_v \left(\frac{g_\omega}{3m_\omega^2} + \frac{g_\rho}{m_\rho^2}\right)\frac{1}{v\cdot k + \Delta}\left(f_D + g f_{D_s}\sqrt{\frac{m_{D_s}}{m_D}} \frac{v\cdot p_1}{(v\cdot p_2 + v\cdot k)}\right)\\
                &- g \lambda g_v \sqrt{\frac{m_{D_s}}{m_D}}f_{D_s} \frac{\sqrt{2}g_\Phi}{3m_\Phi^2}\frac{v\cdot p_1}{(v\cdot p_2 + v\cdot k)(v\cdot p_2 + \Delta)}\\
    B_3^{(s,s)} &= \frac{2}{\sqrt{m_D}} f_K \frac{m^2_{K^{\star}}}{g_{K^{\star}}} g_{K^{\star\pm} K^\pm \gamma} (\alpha_1 m_D - \alpha_2 v\cdot p_1)BW_{K^{\star-}}(p_2+k)\\
                &- g_{K^\star} g_{K^{\star\pm} K^\pm \gamma} \frac{f_D}{f_K} \left(1+ g\sqrt{\frac{m_{D_s}}{m_D}}\frac{f_{D_s}}{f_D}\frac{m_D-v\cdot p_2}{v\cdot p_2 + \Delta}\right) BW_{K^{\star+}}(p_1+k)
  \end{split}
\end{align}
\begin{align}
  \begin{split}
    D_1^{(s,s)} &=f_D \frac{1}{v\cdot k + \Delta} \left(1 - m_\Phi^2 BW_{\Phi}(p_1 + p_2)\right)\left(2\lambda^\prime + \frac{1}{\sqrt{2}} \lambda g_v \left(\frac{g_\omega}{3m_\omega^2} + \frac{g_\rho}{m_\rho^2}\right)\right)\\
    D_1^{(d,d)} &= f_D \frac{1}{v\cdot k + \Delta}\left(m_\omega^2 BW_\omega(p_1+p_2) -m_\rho^2 BW_\rho(p_1+p_2) \right)\left(\lambda^\prime + \frac{1}{2\sqrt{2}} \lambda g_v \left(\frac{g_\omega}{3m_\omega^2} + \frac{g_\rho}{m_\rho^2}\right)\right)\\
    D_2^{(s,s)} &=-\frac{f_D}{f_K} g_{K^\star} g_{K^{\star\pm} K^\pm \gamma} \left(BW_{K^{\star+}}(p_1+k) + BW_{K^{\star-}}(p_2+k)\right)\\
    D_3^{(s,s)} &= -\frac{f_D f_{\eta_8}}{2} \frac{m_D^2}{m_D^2 - m_{\eta_8}^2} \frac{m_{K^\star}^2}{g_{K^\star}} g_{K^{\star\pm} K^\pm \gamma}\left(BW_{K^{\star+}}(p_1+k) +  BW_{K^{\star-}}(p_2+k)\right) \\ 
    D_3^{(d,d)} &= -\frac{f_D}{2} m_D^2 \left(\frac{f_{\eta_8}}{m_D^2 - m_{\eta_8}^2} + \frac{f_{\pi}}{m_D^2 - m_{\pi}^2}\right) \frac{m_{K^\star}^2}{g_{K^\star}} g_{K^{\star\pm} K^\pm \gamma}\left(BW_{K^{\star+}}(p_1+k) +  BW_{K^{\star-}}(p_2+k)\right)
  \end{split}
\end{align}
\begin{align}
  \begin{split}
    &BW_n(x) = \frac{1}{x^2 - m_n^2 + \im m_n \Gamma_n}\, ,\\
    &\Delta = m_{D^\star}-m_D
  \end{split}
\end{align}

\subsubsection{Tensor form factors}

\begin{figure}
  \centering
  \includegraphics[width=0.7\linewidth]{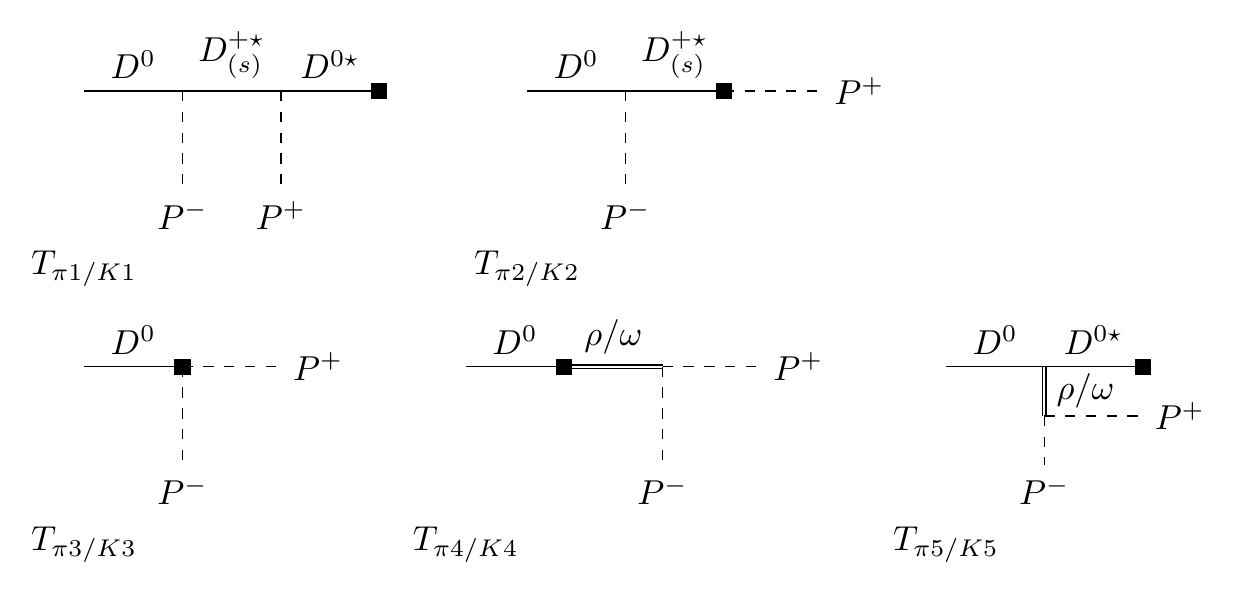}
  \caption{Feynman diagrams contributing to the tensor current form factors $a^\prime$, $b^\prime$, $c^\prime$ and $h^\prime$.}
  \label{fig:Diagramme_Tensorstrom}
\end{figure}

\noindent\underline{$D \to \pi^+ \pi^-$}

\begin{align}
  \begin{split}
    a^\prime &= -\frac{g^2f_D\left(p_2\cdot k -(v\cdot k)(v\cdot p_2)\right)}{f_\pi^2 (v\cdot p_2 + \Delta)(v\cdot p_1 + v\cdot p_2 + \Delta)} + \frac{\alpha_1(v\cdot k)}{f_\pi^2 \sqrt{m_D}}\left[1 + \frac{f_\pi^2 m_\rho^4}{g_\rho^2} BW_\rho(p_1+p_2) \right] \\
    &- {\frac{\sqrt{2}\lambda f_D g_v }{(v\cdot p_1+v\cdot p_2+\Delta)}\frac{m_\rho^2}{g_\rho}\left(p_2\cdot k -(v\cdot k)(v\cdot p_2)\right)BW_\rho(p_1+p_2)}
  \end{split}
\end{align}
\begin{align}
  \begin{split}
    b^\prime &= \frac{gf_D}{f_\pi^2 (v\cdot p_2 + \Delta)}\left[v\cdot k +\frac{g\left(p_1 \cdot k -(v\cdot k)(v\cdot p_1)\right)}{(v\cdot p_1+v\cdot p_2+\Delta)}\right] - {\frac{\alpha_1(v\cdot k)}{f_\pi^2 \sqrt{m_D}}}\left[{1} + {\frac{f_\pi^2 m_\rho^4}{g_\rho^2} BW_\rho(p_1+p_2)}\right] \\
    &+ {\frac{\sqrt{2}\lambda f_D g_v}{(v\cdot p_1+v\cdot p_2+\Delta)}\frac{m_\rho^2}{g_\rho}\left(p_1\cdot k - (v\cdot k)(v\cdot p_1)\right)BW_\rho(p_1+p_2)}
  \end{split}
\end{align}
\begin{align}
  \begin{split}
    c^\prime &= \frac{gf_D}{f_\pi^2 m_D (v\cdot p_2 + \Delta)} \left[-p_2\cdot k  + \frac{g\left((p_2\cdot k)(v\cdot p_1)-(p_1\cdot k)(v\cdot p_2)\right)}{(v\cdot p_1+v\cdot p_2+\Delta)}\right]\\
    &+ {\frac{\alpha_1}{f_\pi^2 m_D^{\frac{3}{2}}}(p_2\cdot k-p_1\cdot k)}\left[{1} + {\frac{f_\pi^2 m_\rho^4}{g_\rho^2} BW_\rho(p_1+p_2)}\right] \\
    &- {\frac{\sqrt{2}\lambda f_D g_v}{m_D(v\cdot p_1+v\cdot p_2+\Delta)}\frac{m_\rho^2}{g_\rho}\left((p_1\cdot k)(v\cdot p_2) - (p_2\cdot k)(v\cdot p_1)\right)BW_\rho(p_1+p_2)}
  \end{split}
\end{align}
\begin{align}
  \begin{split}
    h^\prime &= \frac{gf_D}{2f_\pi^2 m_D (v\cdot p_2 + \Delta)} \left[1 + \frac{g v\cdot k}{(v\cdot p_1+v\cdot p_2+\Delta)}\right] + {\frac{\alpha_1}{f_\pi^2 m_D^{\frac{3}{2}}}}\left[{1} + {\frac{f_\pi^2 m_\rho^4}{g_\rho^2} BW_\rho(p_1+p_2)}\right] \\
    &+ {\frac{\lambda f_D g_v\left(v\cdot k\right)}{\sqrt{2}m_D(v\cdot p_1+v\cdot p_2+\Delta)}\frac{m_\rho^2}{g_\rho}BW_\rho(p_1+p_2)}
  \end{split}
\end{align}

\noindent\underline{$D \to K^+ K^-$}
\begin{align}
  \begin{split}
    a^\prime &= -\frac{g^2f_D\left(p_2\cdot k -(v\cdot k)(v\cdot p_2)\right)}{f_K^2 (v\cdot p_2 + \Delta)(v\cdot p_1 + v\cdot p_2 + \Delta)} \\
    &+ {\frac{\alpha_1(v\cdot k)}{f_K^2 \sqrt{m_D}}}\left[{1} + {\frac{f_K^2}{2} \left(\frac{m_\rho^4}{g_\rho^2}BW_\rho(p_1+p_2) + \frac{m_\omega^4}{g_\omega^2}BW_\omega(p_1+p_2)\right) }\right] \\
    &- {\frac{\lambda f_D g_v\left(p_2\cdot k -(v\cdot k)(v\cdot p_2)\right)}{\sqrt{2}(v\cdot p_1+v\cdot p_2+\Delta)}\left(\frac{m_\rho^2}{g_\rho}BW_\rho(p_1+p_2) + \frac{m_\omega^2}{g_\omega}BW_\omega(p_1+p_2)\right)}
  \end{split}
\end{align}
\begin{align}
  \begin{split}
    b^\prime &= \frac{gf_D}{f_K^2 (v\cdot p_2 + \Delta)}\left[\frac{f_{D_s} \sqrt{m_{D_s}}}{f_D \sqrt{m_D}}v\cdot k +\frac{g\left(p_1 \cdot k -(v\cdot k)(v\cdot p_1)\right)}{(v\cdot p_1+v\cdot p_2+\Delta)}\right] \\
    &- {\frac{\alpha_1(v\cdot k)}{f_K^2 \sqrt{m_D}}}\left[{1} + {\frac{f_K^2}{2} \left(\frac{m_\rho^4}{g_\rho^2}BW_\rho(p_1+p_2) + \frac{m_\omega^4}{g_\omega^2}BW_\omega(p_1+p_2)\right)}\right] \\
    &+ {\frac{\lambda f_D g_v\left(p_1\cdot k - (v\cdot k)(v\cdot p_1)\right)}{\sqrt{2}(v\cdot p_1+v\cdot p_2+\Delta)}\left(\frac{m_\rho^2}{g_\rho}BW_\rho(p_1+p_2) + \frac{m_\omega^2}{g_\omega}BW_\omega(p_1+p_2)\right)}
  \end{split}
\end{align}
\begin{align}
  \begin{split}
    c^\prime &= \frac{gf_D}{f_K^2 m_D (v\cdot p_2 + \Delta)} \left[-\frac{f_{D_s} \sqrt{m_{D_s}}}{f_D \sqrt{m_D}}p_2\cdot k  + \frac{g\left((p_2\cdot k)(v\cdot p_1)-(p_1\cdot k)(v\cdot p_2)\right)}{(v\cdot p_1+v\cdot p_2+\Delta)}\right]\\
    &+ {\frac{\alpha_1}{f_K^2 m_D^{\frac{3}{2}}}(p_2\cdot k-p_1\cdot k)}\left[{1} + {\frac{f_K^2}{2} \left(\frac{m_\rho^4}{g_\rho^2}BW_\rho(p_1+p_2) + \frac{m_\omega^4}{g_\omega^2}BW_\omega(p_1+p_2)\right)}\right] \\
    &- {\frac{\lambda f_D g_v\left((p_1\cdot k)(v\cdot p_2) - (p_2\cdot k)(v\cdot p_1)\right)}{\sqrt{2}m_D(v\cdot p_1+v\cdot p_2+\Delta)}\left(\frac{m_\rho^2}{g_\rho}BW_\rho(p_1+p_2) + \frac{m_\omega^2}{g_\omega}BW_\omega(p_1+p_2)\right)}
  \end{split}
\end{align}
\begin{align}
  \begin{split}
    h^\prime &= \frac{gf_D}{2f_K^2 m_D (v\cdot p_2 + \Delta)} \left[\frac{f_{D_s} \sqrt{m_{D_s}}}{f_D \sqrt{m_D}} + \frac{g v\cdot k}{(v\cdot p_1+v\cdot p_2+\Delta)}\right] \\
    &+ {\frac{\alpha_1}{f_K^2 m_D^{\frac{3}{2}}}}\left[{1} + {\frac{f_K^2}{2} \left(\frac{m_\rho^4}{g_\rho^2}BW_\rho(p_1+p_2) + \frac{m_\omega^4}{g_\omega^2}BW_\omega(p_1+p_2)\right)}\right] \\
    &+ {\frac{\lambda f_D g_v\left(v\cdot k\right)}{2\sqrt{2}m_D(v\cdot p_1+v\cdot p_2+\Delta)}\left(\frac{m_\rho^2}{g_\rho}BW_\rho(p_1+p_2) + \frac{m_\omega^2}{g_\omega}BW_\omega(p_1+p_2)\right)}
  \end{split}
\end{align}

\subsubsection{Differences with respect to \cite{Fajfer:2002bq}}

In the following, we list some differences between our results and
those obtained in Ref.~\cite{Fajfer:2002bq}. Equation numbers refer to
Ref.~\cite{Fajfer:2002bq}.
\begin{enumerate}
	\item Eq (9): the factor $i$ should be absent
	\item Eq (15): the electromagnetic coupling $e$ is missing
	\item Eq (18): the factor $i$ in front of the $A_2$ term is missing
	\item Eq (21): the sign in front of $a$ should be a $+$ (as written in \cite{Bajc:1994ui})
	\item The Wilson coefficients $a_1$ and $a_2$ are missing in the amplitudes in Eqs.~(24) and~(25).
	\item The contributions of the diagrams $A_{4,1}^+$, $C_{4,1}^+$ and $A_{4,1}^0$ vanish in our calculation.
	\item We believe that there are diagrams that have not been shown in Ref.~ \cite{Fajfer:2002bq}: For each of the diagrams $A_{1,1}^0$, $A_{1,2}^0$, $A_{1,3}^0$, $A_{2,2}^0$, $A_{2,3}^0$, $A_{2,4}^0$, $C_{1,1}^0$, $C_{1,2}^0$, $C_{1,3}^0$ and $C_{1,4}^0$ there is another one in which the photon couples via a vector meson. Moreover, we find two additional diagrams for $C^0$. The first one is the same diagram as $A_{2,2}^0$, but with a different factorization. The second is another diagram with a $V \to P P \gamma$ vertex. Only with these two additional diagrams we obtain an expression that is gauge invariant for any value of $a$. However, we obtain $C^0=0$, as in Ref.~\cite{Fajfer:2002bq}.
	\item We reproduce $A_1^+$, but for $A_1^0$ we get an expression $\sim (q\cdot k)-M(v \cdot k + v \cdot q)$.
	\item We have an extra factor of $2$ in $D_3^0$.
	\item We obtain a relative minus sign for each vector meson in a diagram; however, we get the same relative signs for $R_\gamma^{0/+}$ as given in Eqs.~(24) and~(25)~\cite{Bajc:1994ui}.
\end{enumerate}

\end{document}